\documentclass[12pt]{article}

\pdfoutput=1
\usepackage[latin9]{inputenc}
\usepackage[top=80pt,bottom=85pt,left=67pt,right=67pt]{geometry}
\usepackage{amssymb,amsmath}
\usepackage{xypic,subfigure}
\usepackage{graphicx,color}
\usepackage{bbm}
\usepackage{float}
\usepackage{cite}
\usepackage[debug,pageanchor=false]{hyperref}
\definecolor{link}{rgb}{.8,.15,.1}
\hypersetup{colorlinks=true,linkcolor=link,citecolor=link,urlcolor=link,linktocpage}
\DeclareMathOperator{\sech}{sech}

\makeatletter
\@addtoreset{equation}{section}
\makeatother

\newcommand{\beq}{\begin{equation}}
\newcommand{\eeq}{\end{equation}}
\newcommand{\bea}{\begin{eqnarray}}
\newcommand{\eea}{\end{eqnarray}}
\newcommand{\nn}{\nonumber}

\newcommand{\eq}{\begin{equation}}
\newcommand{\feq}{\end{equation}}
\newcommand{\eqn}{\begin{eqnarray}}
\newcommand{\feqn}{\end{eqnarray}}

\begin{document}
\begin{titlepage}

\begin{center}

\vskip .5in 
\noindent

{\Large \bf{New $\text{AdS}_2/\text{CFT}_1$ pairs from $\text{AdS}_3$ and monopole bubbling}}

\bigskip\medskip

Andrea Conti\footnote{contiandrea@uniovi.es}, Yolanda Lozano\footnote{ylozano@uniovi.es},  Niall T. Macpherson\footnote{macphersonniall@uniovi.es}  \\

\bigskip\medskip
{\small 

Department of Physics, University of Oviedo,
Avda. Federico Garcia Lorca s/n, 33007 Oviedo}

\medskip
{\small and}

\medskip
{\small 

Instituto Universitario de Ciencias y Tecnolog\'ias Espaciales de Asturias (ICTEA),\\
Calle de la Independencia 13, 33004 Oviedo, Spain}

\vskip 2cm 

     	{\bf Abstract }
     	\end{center}
     	\noindent
	
We present general results on generating $\text{AdS}_2$ solutions to Type II supergravity from $\text{AdS}_3$ solutions via U(1) and SL(2) T-dualities. We focus on a class of Type IIB solutions with small $\mathcal{N}=4$ supersymmetry, that we show can be embedded into a more general class of solutions obtained by double analytical continuation from $\text{AdS}_3$ geometries with small $\mathcal{N}=(0,4)$ supersymmetry constructed in the literature. We then start the analysis of the superconformal quantum mechanics dual to the $\mathcal{N}=4$ backgrounds focusing on a subclass of $\text{AdS}_2\times\text{S}^3\times\mathbb{T}^3$ solutions foliated over a Riemann surface.  We show that the associated supersymmetric quantum mechanics describes monopole bubbling in 4d $\mathcal{N}=2$ supersymmetric gauge theories living in D3-D7 branes, as previously discussed in the literature. Therefore, we propose that our solutions provide a geometrical description via holography of monopole bubbling in 4d $\mathcal{N}=2$ SCFTs. We check our proposal with the computation of the central charge. 
    	
\noindent

\vfill
\eject

\end{titlepage}

\tableofcontents

\section{Introduction}

The $\text{AdS}_2/\text{CFT}_1$ correspondence plays a key role in the microscopical description of extremal black holes, $\text{AdS}_2$ being part of the  geometry that appears in their near horizon limit in any dimension. It presents however important conceptual problems that have to do mainly with the non-connected nature of the $\text{AdS}_2$ boundary and the fact that $\text{AdS}_2$ gravity is non-dynamical.

In recent years new examples of $\text{AdS}_2$ geometries with different supersymmetries have been constructed \cite{Lozano:2020txg,Lozano:2020sae,Lozano:2021rmk,Ramirez:2021tkd,Lozano:2021xxs,Lozano:2021fkk,Lozano:2022vsv,Lozano:2022swp} (see also  \cite{Dibitetto:2019nyz,Aniceto:2020saj}) that show that when extra fluxes are added $\text{AdS}_2$ gravity ceases to be non-dynamical, and allows for a non-vanishing central extension from which the black hole entropy can be computed. These examples extend the $\text{AdS}_2$ geometries with constant electric fields proposed early on  in \cite{Strominger:1998yg,Balasubramanian:2003kq,Hartman:2008dq,Alishahiha:2008tv,Balasubramanian:2009bg}. In some of these constructions the dual superconformal quantum mechanics (SCQM) was also identified, and it was possible to check the conjecture in  \cite{Hartman:2008dq}, namely, that a consistent quantum gravity on $\text{AdS}_2$ should be dual to a chiral half 2d CFT,  with explicit realisations. Indeed, in these constructions the supersymmetric quantum mechanics from which the SCQMs arise in the IR are compactifications of 2d (0,4) SCFTs, and have the same superconformal algebras associated to them. This has been used to compute the central charge of the SCQMs, and perfect agreement has been found with the corresponding holographic expressions. The proposed SCQMs thus provide  explicit scenarios where the entropy of extremal black holes can be computed microscopically.

Another useful application of the $\text{AdS}_2/\text{CFT}_1$ correspondence is to the holographic description of superconformal line defects in higher dimensional CFTs \cite{Dibitetto:2018gtk,Gutperle:2018fea,Chen:2019qib,Chen:2020mtv,Lozano:2021fkk,Lozano:2022vsv,Lozano:2022swp}. Geometrically, a sign that an $\text{AdS}_2$ solution may be describing a superconformal line defect is that it flows asymptotically locally to a higher dimensional AdS background, dual far from the defects to the higher dimensional CFT in which they are embedded. In some of these constructions it has been shown that the quiver quantum mechanics dual to the $\text{AdS}_2$ solutions can be embedded within the quivers that describe the higher dimensional SCFTs \cite{Lozano:2021fkk,Lozano:2022vsv}. In some instances the defects are described by quivers living in brane boxes \cite{Lozano:2022vsv}, with the latter finding in this way a powerful holographic realisation. 

In this paper we present general results on the construction of $\text{AdS}_2$ solutions to Type II supergravity via U(1) and SL(2) T-dualities. We then exploit these to construct new  classes of small $\mathcal{N}=4$ solutions in Type II supergravity and study the dual field theory description of a subclass of these solutions, for which the SCQM arises once more upon compactification of a 2d (0,4) SCFT. 
The intricacy of the brane set-up associated to these
solutions reveals an interesting defect interpretation in terms of baryon vertices and 't Hooft loops within higher dimensional CFTs.

The paper is organised as follows. We start in section 2 presenting general results on the generation of $\text{AdS}_2$ solutions in Type II supergravity from $\text{AdS}_3$ solutions via U(1) and SL(2) T-dualities, paying special attention to the conditions for preservation of supersymmetry. This section is supplemented by appendices  \ref{sec:appendixonconvensions}-\ref{sec:appendixonSUSY}. In section 3 we focus on the particular classes of solutions to Type IIB supergravity preserving $\mathcal{N}=4$ supersymmetry obtained from the $\text{AdS}_3\times \text{S}^3\times \text{M}_4$ solutions to massive IIA with $\mathcal{N}=(0,4)$ supersymmetry and SU(3) structure constructed in \cite{Lozano:2022ouq}. We show that the solutions with a 3-torus isometry can be embedded into a more general class of solutions, that we construct via double analytical continuation from the $\text{AdS}_3$ solutions in Type IIB with $\mathcal{N}=(0,4)$ supersymmetry constructed in \cite{Macpherson:2022sbs}. Then in section \ref{field-theory} we move to the field theory description of a subclass of the previous solutions, consisting on $\text{AdS}_2\times \text{S}^3\times \mathbb{T}^3$ foliations over a 2d Riemann surface, parametrised by two non-compact directions $(\rho,r)$. 
We show that the associated quantum mechanics lives in D1-branes stretched between NS5-branes along the $\rho$ direction, with D3, D5 and D7 branes contributing with flavour groups to the resulting (0,4) quantum mechanics, that, we propose,  flows in the IR to the SCQMs dual to the solutions. Making contact with similar 1d quivers proposed in the literature in the description of bubbling dyonic monopoles in 4d $\mathcal{N}=2$ supersymmetric field theories living in D3-D7 systems \cite{Brennan:2018yuj,Brennan:2018moe,Brennan:2018rcn,Assel:2019iae}, we identify our quiver constructions as describing the  bubbling sector of vanishing effective magnetic charge in these 4d theories. In turn, the disposition of the branes along the $r$ direction allows us to interpret the F1-strings present in the brane set-up in relation to baryon vertices in the 5d theory living in the D5-NS5-D7 subset of branes. 
We check our holographic proposal with the computation of the central charge on both sides of the AdS/CFT correspondence. 
On the field theory side we compute the central charge from the R-symmetry anomaly, using the same expression valid in two dimensions. We find perfect agreement with the holographic result. Appendix \ref{fieldtheory} contains technical details of the construction of the 1d quivers discussed in section \ref{field-theory}.

\section{Generating AdS$_2$ solutions from U(1) and SL(2) T-duality on AdS$_3$}\label{generalsols}

In this section we present some general results on generating AdS$_2$ solutions of Type II supergravity from AdS$_3$ solutions via U(1) and SL(2) T-dualities. In particular we present the general form of such solutions and explain how they preserve supersymmetry. The content of this section is extensively supplemented by the technical appendices \ref{sec:appendixonconvensions}-\ref{sec:appendixonSUSY} where the results are derived.\\
~\\
An AdS$_3$ solution of Type II  supergravity can in general be decomposed in the form 
\begin{align}
\label{eq:1.0}
ds^2&=e^{2A}ds^2(\text{AdS}_3)+ ds^2(\text{M}_7),\nn\\[2mm]
H^{(10)}&=c_0\text{vol}(\text{AdS}_3)+H,~~~~ F= f_{\pm}+ e^{3A}\text{vol}(\text{AdS}_3)\wedge\star_7 \lambda  f_{\pm},
\end{align}
where $(e^{2A},f_{\pm},H)$ and the dilaton $\Phi$ have support on $\text{M}_7$, $c_0$ is a constant and the upper/lower signs are taken in IIA/IIB. We shall be interested in a subset of these for which 
\beq
c_0=0,~~~~ H^{(10)}=H,
\eeq
ie those with purely magnetic NS 3-form flux, otherwise the potential for $H^{(10)}$ will not be SL(2) invariant and we will be unable to perform a non-Abelian T-duality transformation on it\footnote{Note that an electric component of the NS flux does not pose a problem for U(1) T-duality on the Hopf fiber of AdS$_3$, however the dual solution would not have a round AdS$_2$ factor - this would instead appear with a U(1) fibred over it in the dual solution.}. The Bianchi identities the fluxes  should obey (away form the loci of sources)  reduce to the $d=7$ conditions
\beq
\label{eq:1.1}
dH=0,~~~ (d-H\wedge) f_{\pm}=0,~~~~ (d-H\wedge)(e^{3A}\star_7 \lambda (f_{\pm}))=0.
\eeq
When an AdS$_3$ solution preserves at least ${\cal N}=1$ supersymmetry it admits at least a pair of  Majorana spinors $(\chi_1,\chi_2)$ on M$_7$ such that one can define a pair of $d=10$ Majorana-Weyl spinors decomping as
\beq
\epsilon_1=  \zeta\otimes \theta_+\otimes \chi_1,~~~~~\epsilon_2= \zeta\otimes \theta_{\mp}\otimes \chi_2,\label{eq:susydecompostions}
\eeq
where $\zeta$ is a Killing spinor on unit radius AdS$_3$ obeying the equation
\beq
\nabla^{\text{AdS}_3}_{\mu}\zeta=  \frac{s}{2}\gamma_{\mu}\zeta,~~~\mu=0,1,2~~~~s^2=1.\label{eq:AdS3KSEmain}
\eeq
For such spinors the necessary conditions for supersymmetry ultimately reduce to a set of $d=7$ conditions involving only $(\chi_1,\chi_2)$ (see   \eqref{seedsolutionsusyeqs1}-\eqref{seedsolutionsusyeqs6}).  The specific sign $s$ takes is related to the type of chiral algebra one has on the boundary of AdS$_3$, ${\cal N}=(1,0)$ or ${\cal N}=(0,1)$.  Which corresponds to which sign is a matter of convention and the literature has no set standard.  The remaining terms $\theta_{\pm}$ are 2 dimensional vectors which account for $d=10$ chirality, again the upper/lower signs are taken in Type IIA/IIB - further details can be found in appendix \ref{eq:AdS3SUSYapppendix}. Of course a generic AdS$_3$ solution can preserve ${\cal N}=(p,q)$ supersymmetry for $p,q$ integers such that $p+q\leq 8$ \cite{Haupt:2018gap}. In this case one will actually have $p$ independent versions of \eqref{eq:susydecompostions} coupled to Killing spinors on AdS$_3$ that obey  \eqref{eq:AdS3KSEmain} for one sign and a further $q$ versions with AdS$_3$ Killing spinors obeying the opposite sign - each of these comes equipped with an independent $(\chi_1,\chi_2)$ pair.\\
~\\
It is possible to generate an AdS$_2$ solution from an AdS$_3$ background by performing a U(1) T-duality on the Hopf fiber of AdS$_3$. The result of doing this to \eqref{eq:1.0} with $c_0=0$ is the following 
\begin{align}
ds^2_{\mathfrak{u}(1)}&= \frac{e^{2A}}{4} ds^2(\text{AdS}_2)+ e^{-2A} dr^2+ ds^2(\text{M}_7),~~~~e^{- \Phi_{\mathfrak{u}(1)}}= e^{-\Phi+A},\nn\\[2mm]
 H_{\mathfrak{u}(1)}&=  -\frac{s}{2}r\text{vol}(\text{AdS}_2)+H,~~~~F_{\mathfrak{u}(1)}=t (f_{\pm}\wedge dr\mp \frac{1}{4}e^{3A}\text{vol}(\text{AdS}_2)\wedge \star_7 \lambda f_{\pm}), \label{eq:U1Tdual}
\end{align}
where $t^2=1$ and the upper/lower signs refer to whether the original solution was in IIA/IIB, with the dual solution presented above in IIB/IIA.  The vector $\partial_r$ is Killing with respect to the entire solution, and if M$_7$ was bounded the generated AdS$_2$ solution will be likewise bounded if $r$ is periodically identified. It is a simple matter to show that the Bianchi identities of the the dual flux 
\beq
(d- H_{\mathfrak{u}(1)}\wedge )F_{\mathfrak{u}(1)}=0,~~~ d H_{\mathfrak{u}(1)}=0,
\eeq
are implied by \eqref{eq:1.1}.  Likewise the remaining equations of motion of the bosonic supergravity fields \eqref{eq:EOM} are implied by those following from \eqref{eq:1.0}. 

If the original solution was supersymmetric the dual solution preserves all of the chiral supercharges of a single chirality, determined by the choice $s=1$ or $s=-1$ so if the original solution was ${\cal N}=(p,q)$ supersymmetric the dual can preserve one of ${\cal N}=p$ or ${\cal N}=q$ supersymmetry. The dual  $d=10$ Killing spinors share the same internal spinors as \eqref{eq:susydecompostions}, they take the form
\beq
\epsilon^{\mathfrak{u}(1)}_1=  \zeta_2\otimes \theta_+\otimes \chi_1,~~~~\epsilon_2^{\mathfrak{u}(1)}= -t\gamma_{\underline{r}}\zeta_2\otimes \theta_{\pm}\otimes \chi_2,\label{eq:U(1)susydecompostions}
\eeq
where $\zeta_2$ are Killing spinors on unit radius AdS$_2$ obeying
\beq
\nabla^{\text{AdS}_2}_{\mu}\zeta_2=  \frac{s}{2}\gamma_{\mu}\zeta_2,~~~\mu=0,1~~~~s^2=1\label{eq:AdS2KSEmain},
\eeq 
and there exists frames where $\zeta=\zeta_2$.\\
~\\
Generating AdS$_2$ solutions from AdS$_3$ with Hopf fiber T-duality is an old idea, a more novel way to generate AdS$_2$ solutions is to instead utilise SL(2) non-Abelian T-duality, as recently done in 
\cite{Lozano:2021rmk,Ramirez:2021tkd}. For completeness
we spell out the process of how this is done in appendix \ref{sec:performingSL(2)T-duality}. The result of applying an SL(2) T-duality on \eqref{eq:1.0} with $c_0=0$ is the following solution in Type IIB/IIA
\begin{align}
ds^2_{\mathfrak{sl}(2)}&= \frac{e^{2A}}{4\Delta} ds^2(\text{AdS}_2)+ e^{-2A} dr^2+ ds^2(\text{M}_7),~~~~e^{- \Phi_{\mathfrak{sl}(2)}}= e^{-\Phi+A}r\sqrt{\Delta} ,\nn\\[2mm]
 H_{\mathfrak{sl}(2)}&=  -\frac{s}{2\Delta}r\text{vol}(\text{AdS}_2)+H,~~~~~\Delta= 1-\frac{e^{4A}}{4r^2}\nn\\[2mm]
 F_{\mathfrak{sl}(2)}&= t\bigg[f_{\pm}\wedge( r -\frac{s e^{4A}}{8\Delta}\text{vol}(\text{AdS}_2))\wedge dr\mp\frac{e^{3A}}{2}(-s+ \frac{r}{2\Delta} \text{vol}(\text{AdS}_2))\wedge \star_7 \lambda f_{\pm}\bigg], \label{eq:SL2Tdual}
\end{align}
where again $t^2=1$. Unlike the case of U(1) T-duality, $\partial_r$ is not an isometry of \eqref{eq:SL2Tdual} and cannot be periodically identified. In fact as with SU(2) T-duality the dual $r$ coordinate is now a semi-infinite interval:  As $r\to \infty$ the metric exhibits no regular zeros or singularites so $r$ is not bounded from above, conversely as $r$ approaches loci for which $\Delta=0$ the behaviour of OF1 planes  (which are the S-dual of O1 planes) is recovered \cite{Ramirez:2021tkd}.  Again one can show that the Bianchi identities of the fluxes 
\beq
(d- H_{\mathfrak{sl}(2)}\wedge )F_{\mathfrak{sl}(2)}=0,~~~ d H_{\mathfrak{sl}(2)}=0,
\eeq
are implied by \eqref{eq:1.1}, and  that the rest of the Type II equations of motion \eqref{eq:EOM} are implied by the reduced $d=7$ conditions following from \eqref{eq:1.0}. 

An interesting fact emerges while comparing  \eqref{eq:SL2Tdual} to \eqref{eq:U1Tdual}, given that $\Delta \to 1$ as $r\to \infty$. It quickly becomes clear that the NS sector of the U(1) T-dual is recovered from that of the SL(2) T-dual as
\beq
\lim_{r\to \infty}(ds^2_{\mathfrak{sl}(2)},~ H_{\mathfrak{sl}(2)},~r e^{\Phi_{\mathfrak{sl}(2)}})= (ds^2_{\mathfrak{u}(1)},~ H_{\mathfrak{u}(1)},~e^{\Phi_{\mathfrak{u}(1)}}).
\eeq
A similar map can be obtained for the RR sector, when weighted by the dilaton, ie
\beq
\lim_{r\to \infty}e^{\Phi_{\mathfrak{sl}(2)}} F_{\mathfrak{sl}(2)}=e^{\Phi_{\mathfrak{u}(1)}}F_{\mathfrak{u}(1)},
\eeq
holds in general. Similar observations have been made for SU(2) T-duality \cite{Lozano:2016wrs}.

If the original AdS$_3$ solution is supersymmetric, like the U(1) case, the dual solution can preserve all the spinors corresponding to a given chirality determined by the sign of $s$. The dual Killing spinors take the form  
\begin{align}
\epsilon^{\mathfrak{sl}(2)}_1&= (N_++N_-\gamma_{\underline{r}})\zeta_2\otimes \theta_{+}\otimes \chi_1,~~~~\epsilon^{\mathfrak{sl}(2)}_2=-t(N_+-N_-\gamma_{\underline{r}})\gamma_{\underline{r}}\zeta_2\otimes \theta_{\pm}\otimes \chi_2,\nn\\[2mm]
N_{\pm}&=\frac{1}{2\Delta^{\frac{1}{4}}}(\sqrt{\Delta_+}\pm \sqrt{\Delta_-}),~~~~\Delta_{\pm}=1\pm\frac{e^{2A}s}{2r},
\end{align}
where again $\zeta_2$ are Killing spinors on AdS$_2$ obeying \eqref{eq:AdS2KSEmain}.\\
~\\
In the next section we shall utilise the technology presented in this section to generate a new class of AdS$_2$ solutions in Type II supergravity preserving small ${\cal N}=4$ supersymmetry.

\section{New classes of small ${\cal N}=4$ AdS$_2$ solutions in Type IIB }

In section \ref{sec:newclass} we derive a new class of small ${\cal N}=4$ AdS$_2$ solutions of Type II supergravity via SL(2) T-duality. In section \ref{sec:twosolutions} we present two solutions within this class whose geometries are warped products of $\text{AdS}_2\times \text{S}^3\times \mathbb{T}^3\times \Sigma_2$. We then derive a general class of solutions with this topology in section \ref{sec:genclass} with a view towards ``completing'' the SL(2) T-duals within global solutions with bounded internal space in the future. As a first step we confirm that they do indeed lie within this broader class.

\subsection{ A small ${\cal N}=(0,4)$ AdS$_3$ class in IIA and its SL(2) T-duality}\label{sec:newclass}

We are interested in studying the SL(2) T-duality of a class of small ${\cal N}=(0,4)$ AdS$_3$ solutions in massive IIA first derived in \cite{Lozano:2022ouq}. The general class has a NS sector of the form
\begin{align}
\label{eq:2.0}
ds^2 & =  \frac{q}{\sqrt{h}} \left( ds^2(\text{AdS}_3) + ds^2(\text{S}^3) \right) + g \sqrt{h} dx_i^2  + \frac{g d\rho^2 }{\sqrt{h}}, \qquad e^{-\Phi} =\frac{h^{\frac{3}{4}}}{\sqrt{g}}, \nn \\
H & = \partial_{\rho}(gh) dx_1\wedge dx_2\wedge dx_3- \frac{1}{2}\epsilon_{ijk} \partial_{x_i}g dx_j \wedge dx_k \wedge d\rho,
\end{align}
where $i=1,2,3$ with Einstein summation conventions assumed, $g=g(x_i)$, $h=h(\rho,x_i)$, $q$ is a constant and $\epsilon_{ijk}$ is the flat space Levi-Civita symbol. All possible $d=10$ RR fluxes are non trivial and take the form
\begin{align}
\label{eq:2.1}
F_0 &= \frac{ \partial_{\rho} h}{g},~~~F_2 = -\frac{1}{2} \epsilon_{ijk}\partial_{x_i} h dx_j \wedge dx_k,~~~F_4 = 2 q  \left(\text{vol}(\text{AdS}_3)+ \text{vol}(\text{S}^3)\right)\wedge d\rho.
\end{align}
The Bianchi identities of the NS and RR fluxes, away from the loci of sources require that we impose that
\beq  
\label{eq:2.5}
d(\frac{ \partial_{\rho} h}{g})=0,~~~~ \partial_{x_i}^2 g + \partial_{\rho}^2 (g h) =0,~~~~\partial_{x_i}^2 h  + F_0  \partial_{\rho} (gh) =0.
\eeq
If these requirements are satisfied then \eqref{eq:2.0} and \eqref{eq:2.1} always yield a solution of the full Type IIA equations of motion preserving at least ${\cal N}=(0,4)$, however this is actually enhanced to (4,4)  when $h=\text{constant}$.

With the recipe presented in the previous section it is a simple matter to compute the SL(2) T-dual of the above class of solutions, we need only put it in the form of \eqref{eq:1.0} to read off the answer. We have 
\begin{align}
e^A&=\frac{\sqrt{q}}{h^{\frac{1}{4}}},~~~~ ds^2(\text{M}_7)= \frac{q}{\sqrt{h}} ds^2(\text{S}^3) +g \left(\sqrt{h}  dx_i^2 + \frac{ d\rho^2 }{\sqrt{h}}\right), \qquad e^{-\Phi} =\frac{h^{\frac{3}{4}}}{\sqrt{g}},\nn\\[2mm]
f_+&=\frac{ \partial_{\rho} h}{g}-\frac{1}{2}\epsilon_{ijk} \partial_{x_i }hdx_j\wedge dx_k+2 q   \text{vol}(\text{S}^3)\wedge d\rho +2 q g h dx_1 \wedge dx_2 \wedge dx_3 \wedge \text{vol}(\text{S}^3),
\end{align}
with $H$ simply given as in \eqref{eq:2.0}. The sign appearing in the AdS$_3$ Killing spinor equation \eqref{eq:AdS3KSEmain} that \cite{Lozano:2022ouq} takes is
\beq
s=1,
\eeq
so we should choose this sign for small ${\cal N}=4$ to be preserved in the dual solution. Just to make a concrete choice we additionally fix $t=-1$.\\
~\\
The SL(2) T-dual solution in Type IIB supergravity then has the following NS sector
\begin{align}
\label{eq:dualingeneral}
ds^2 & =  \frac{q}{\sqrt{h}} \left( \frac{1}{2\Delta}ds^2(\text{AdS}_2) + ds^2(\text{S}^3) \right)+ \frac{\sqrt{h}}{q} dr^2 + g \sqrt{h}dx_i^2 + \frac{g d\rho^2 }{\sqrt{h}}, \qquad e^{-\Phi} =r\sqrt{\frac{qh\Delta}{g}}, \nn \\
H & = -\frac{1}{2\Delta}r\text{vol}(\text{AdS}_2)+ \partial_{\rho}(gh) dx_1\wedge dx_2\wedge dx_3- \frac{1}{2}\epsilon_{ijk}\partial_{x_i}g dx_j\wedge dx_k\wedge d\rho,
\end{align}
where now $\Delta=1-\frac{q^2}{4 r^2 h}$. The RR sector on the other hand is given by
\begin{align}
F_1&=q d\rho-\frac{\partial_{\rho}h}{g}r dr,~~~F_5= q d\left((1-2\Delta)r^2\right)\wedge \text{vol}(\text{S}^3)\wedge d\rho\nn\\[2mm]
&+\frac{q}{2\Delta}\text{vol}(\text{AdS}_2)\wedge \bigg[r g hdx_1\wedge dx_2\wedge dx_3-\frac{q}{8}\epsilon_{ijk}\partial_{x_i}\log h dr\wedge  dx_j\wedge dx_k\bigg],\\[2mm]
F_3&=-\frac{q}{2\Delta}\text{vol}(\text{AdS}_2)\wedge \left(r d\rho-\frac{q \partial_{\rho}h}{4 g h}dr\right)-q gh dx_1\wedge dx_2\wedge dx_3+\frac{r}{2}\epsilon_{ijk}\partial_{x_i}h dr\wedge  dx_j\wedge dx_k \nn.
\end{align}
As with the original solution, the Bianchi identities and remaining Type II equations of motion hold when \eqref{eq:2.5} or its source corrected equivalent do. 

We will now focus on the SL(2) T-duals of two solutions found and studied in \cite{Lozano:2022ouq} for which concrete CFT duals have been proposed.

\subsection{Two solutions on AdS$_2\times\text{S}^3\times\mathbb{T}^3\times \Sigma_2$}\label{sec:twosolutions}

Two interesting classes of AdS$_2$ solutions arise by imposing  that $x_i$ span a 3-torus which the warp factors respect the isometries of. Thus we restrict to $h=h(\rho)$ and $g=g(\rho)$. One then has that the last of \eqref{eq:2.5} reduces to simply
\beq
F_0\partial_{\rho} (gh)=0,\label{eq:splitting}
\eeq
representing a branching of possible solutions. Either $h \propto g^{-1} $ or $F_0=0$  which requires $h=$ constant and that $g$ is locally a linear function -  we present these respective cases in sections \ref{eq:case1SL2Tdual} and \ref{eq:case2SL2Tdual}.

\subsubsection{AdS$_2$ case with $h g \propto 1$}\label{eq:case1SL2Tdual}
For the first case we solve the whole of  \eqref{eq:2.5} as
\begin{align}
\label{eq:4.4}
h=\sqrt{u}, \qquad g=\frac{c}{\sqrt{u}},~~~~u=u(\rho),
\end{align}
where $u$ is a linear function (ie $u''=0$ locally) and $c$ is a constant. As a consequence we have $\partial_{\rho} (gh)=0$ and so the NS 3-form becomes purely electric.  Globally in massive IIA this tuning of $(h,g)$ gives rise to AdS$_3\times\text{S}^3\times\mathbb{T}^3$ foliated over an interval bounded between D8/O8 singularities and with D8 branes along the interior.

The SL(2) T-dual of the solution described above takes the following form
\begin{align}
ds^2 & = \frac{q}{u^{\frac{1}{4}}}\bigg(\frac{1}{4\Delta_1} ds^2(\text{AdS}_2)+ ds^2(\text{S}^3)+ c ds^2(\mathbb{T}^3)\bigg)+ u^{\frac{1}{4}}\left(\frac{dr^2}{q}+ \frac{c d\rho^2}{u}\right),~~~ e^{-\Phi} =r\sqrt{\frac{q u\Delta_1}{c}}, \nn \\
H & =d B,~~~B=- \frac{r}{2\Delta_1}\text{vol}(\text{AdS}_2) ,~~~~\Delta_1=1- \frac{q^2}{4 r^2 \sqrt{u}}.
\end{align}
As there is a well defined NS 2-form for this solution we find it convenient to express the RR sector in terms of their Page flux avatars defined as
\beq
\hat{F} = e^{-B}\wedge F,\label{eq:standardPagecharges}
\eeq
which are closed if $(d-H\wedge) F=0$. We find the following
\begin{align}
F_1&=q d\rho-\frac{r u'}{2c} dr,~~~\hat{F}_3=-\frac{r^2 u'}{4c}\text{vol}(\text{AdS}_2)\wedge dx-qc \text{vol}(\mathbb{T}^3),\nn~~~
\hat{F}_5=2q r \text{vol}(\text{S}^3)\wedge dr\wedge d\rho,\nn\\[2mm]
\hat F_7&=\text{vol}(\text{S}^3)\wedge\bigg(q r^2\text{vol}(\text{AdS}_2)\wedge dr\wedge d\rho+ cq \text{vol}(\mathbb{T}^3)\wedge \left(2r dr- \frac{q}{2}d(u^{-\frac{1}{2}})\right)\bigg),\nn\\[2mm]
\hat F_9&=cq r^2 \text{vol}(\text{AdS}_2)\wedge\text{vol}(\text{S}^3)\wedge\text{vol}(\mathbb{T}^3)\wedge dr,
\end{align}
where of course $\hat F_1=F_1$. One has a solution away from the loci of sources so long as $u''=0$, making $u$ linear. Globally however $u$ needs only be piecewise linear with the discontinuities in $u'$ giving rise to delta function sources in $u''$, which are D7 branes smeared over $r$. Such objects behave like D8 branes, ie the metric and dilaton neither go to zero nor blow up at their loci, as such they can be placed along the interior of the interval spanned by  $\rho$.

\subsubsection{AdS$_2$ case with $h=$ constant}\label{eq:case2SL2Tdual}

The second way to solve \eqref{eq:splitting} is to fix $F_0=0$ which, given that the solution must also respect the isometries of $\mathbb{T}^3$ means we shall fix
\beq
h= h_0,
\eeq
for $h_0$ a constant.  We then have that \eqref{eq:2.5} reduces to 
\begin{align}\label{eq:gode}
g''=0,
\end{align}
away from the loci of sources. In massless IIA this tuning of $(h,g)$ gives rise to another solution with AdS$_3\times\text{S}^3\times\mathbb{T}^3$ foliated over an interval. This time instead of D8 branes there are NS5 branes, that are smeared over all but one of their co-dimensions, placed along the interior of the interval and giving rise to delta function source corrections to \eqref{eq:gode}.
 
The SL(2) T-dual solution has a NS sector of the following form
\begin{align}
\label{eq:4.1}
ds^2 & = \frac{1}{\sqrt{h_0}}\left(\frac{q}{4\Delta_2}ds^2(\text{AdS}_2)+qds^2(\text{S}^3)\right)+\frac{\sqrt{h_0}}{q}dr^2+ g \left(\sqrt{h_0}ds^2(\mathbb{T}^3)+ \frac{1}{\sqrt{h_0}}d\rho^2\right),\nn\\[2mm]
e^{-\Phi}&=\sqrt{\frac{h_0kq \Delta_2}{g}},~~~ H=-\frac{1}{2}d(\frac{r}{\Delta_2})\wedge \text{vol}(\text{AdS}_2)+h_0 g' \text{vol}(\mathbb{T}^3),~~~~\Delta_2=  1- \frac{q^2}{4h_0 r^2}.
\end{align}
Note that $H$ no longer has a well defined NS 2-form so using \eqref{eq:standardPagecharges} to simplify the expressions of the RR sector is no longer appropriate. The $d=10$ RR fluxes are given by
\begin{align}
F_1&= q d\rho,~~~F_3=-q\left(\frac{ r}{2\Delta_2}\text{vol}(\text{AdS}_2)\wedge d\rho+ h_0 g \text{vol}(\text{T}^3)\right),\nn\\[2mm]
F_5&= qr \left(\frac{h_0 g}{2 \Delta_2}\text{vol}(\text{AdS}_2)\wedge \text{vol}(\mathbb{T}^3)+ 2\text{vol}(\text{S}^3)dr\wedge d\rho \right).\label{eq:4.1b}
\end{align}

In section \ref{field-theory} we study the SCQM dual to this class of solutions, as the starting point for a more general study of SCQMs dual to $\text{AdS}_2$ solutions with $\mathcal{N}=4$ supersymmetries that would include the solutions with $h g \propto 1$ and the broader class that we construct in the next subsection. We argue that the solutions in this subsection describe geometrically the bubbling of monopoles in 4d $\mathcal{N}=2$ D3-D7 theories. 

\subsection{Embedding into a general class of AdS$_2\times\text{S}^3\times\mathbb{T}^3\times \Sigma_2$ solutions}\label{sec:genclass}
An ever present issue with applying  non-Abelian generalisations of U(1) T-duality to the AdS/CFT correspondence is that they give rise to solutions with an unbounded internal space. SL(2) T-duality is no different, the T-dual coordinate $r$ is only bounded at one end. This is often a sign that the putative dual CFT/CQM is ill defined, although it may also indicate that the geometry is dual to a defect in a higher dimensional CFT.  Significant progress on both the geometry and CFT side has been made towards making sense of such geometries in recent years. The CFT dual of a non-Abelian T-dual geometry can often be viewed as an infinite linear quiver that one can make finite, by instead choosing to terminate it at a flavour node  
\cite{Lozano:2016kum,Lozano:2016wrs,Lozano:2017ole,Itsios:2017cew,Lozano:2018pcp,Lozano:2019ywa,Ramirez:2021tkd,Lozano:2022vsv}. To realise such a method geometrically one needs to embed the non-Abelian T-dual geometry into a broader class of solutions and essentially glue another solution onto it at finite $r$, such that the interval spanned by $r$ becomes bounded and the dual CFT well defined. This has been successfully achieved in several contexts. With this in mind in this section we are interested in embedding the solutions of sections \ref{eq:case1SL2Tdual} and \ref{eq:case2SL2Tdual} into a more general class of AdS$_2\times\text{S}^3\times\mathbb{T}^3\times \Sigma_2$  solutions preserving small ${\cal N}=4$ supersymmetry. To our knowledge such a  class does not currently exist, so here we shall derive one by double Wick rotating (a subclass of) a broad class of solutions on $\text{AdS}_3\times \text{S}^2$ found in \cite{Macpherson:2022sbs}.\\
~~\\
The relevant class of AdS$_3\times \text{S}^2\times \text{M}_5$ solutions is presented in section 3.3 of \cite{Macpherson:2022sbs}. The internal manifold M$_5$ is spanned by coordinates $(y,x,z_1,z_2,z_3)$, with the former 2 coordinates fibered over the latter. Generically the class is quite complicated  but undergoes a pronounced simplification if one assumes that  $\partial_{z_{i}}$ are isometries of the solutions spanning a 3-torus. It is then possible to perform a coordinate transformation in the $(x,y)$ directions to  diagonalise the metric. Let us describe the process briefly, note that $(\lambda,g_0,h_0,u_0)$ refer to functions defined in \cite{Macpherson:2022sbs} (the latter 2 not containing the subscript zero in that work). After imposing that $\partial_{z_{i}}$ are isometries one first performs the coordinate transformation
\begin{align}
x \rightarrow \tilde{x}=x, \qquad y \rightarrow 4\tilde{y}= 4 \tilde{y}(y,x) 
\end{align}
in such a way that
\begin{equation}
\lambda = \frac{\partial \tilde{y}}{\partial x}, \qquad g_0 = \frac{\partial \tilde{y}}{\partial y},
\end{equation}
 which leads to  a diagonal metric. One then redefines
\begin{align}
g_0= \frac{4}{K},~~~~ h_0 =4S,~~~~u_0= v, ~~~~ m= 1\label{eq:redfs},
\end{align}
where $K=K(x,y)$, $S=S(x,y)$ and $v=v(x)$. Finally we send $\tilde{y} \rightarrow y$ to uncluttered notation.\\
~~\\
The resulting class of small ${\cal N}=(0,4)$ AdS$_3$ solutions has a NS sector of the following form
\begin{align}
ds^2& = \frac{v}{\sqrt{K S}}\bigg(ds^2(\text{AdS}_3)+\frac{1}{\Xi}ds^2(\text{S}^2)\bigg)+ \frac{\sqrt{SK}}{v} dx^2 +\frac{1}{\sqrt{SK}} dy^2 +\sqrt{\frac{S}{K}} ds^2(\mathbb{T}^3), \nn \\[2mm]
e^{-\Phi} &=  K\sqrt{\frac{ S\Xi}{v}},~~~~ H = \frac{1}{2} d\left(\frac{ v v'}{4 K S \Xi}-x\right)\wedge \text{vol}(\text{S}^2) +\frac{ \partial_y S }{K}  \text{vol}(\mathbb{T}^3),~~~~\Xi=1 + \frac{v'^2}{4K S}.
\end{align}
It supports the following non trivial $d=10$ RR fluxes
\begin{align}
F_1 & = \partial_x K dy -\frac{1 }{v} \partial_y ( K S ) dx,~~~~F_5= (1+\star_{10})f_5 \nn  \\[2mm]
F_3 & =\left(-\frac{ v' \partial_y ( K S )}{8 K S \Xi} dx +\frac{ v v' \partial_x\log K dy }{8S \Xi}-\frac{1}{2} K dy \right)\wedge \text{vol}(\text{S}^2) - \partial_x S \text{vol}(\mathbb{T}^3), \nn \\[2mm]
f_5 & = \frac{1}{ 8 K S \Xi} \left( 4 K S^2-  v^2 v' \partial_x \left(\frac{S}{v}\right) \right) \text{vol}(\mathbb{T}^3) \wedge \text{vol}(\text{AdS}_2).
\end{align}
One has a solution whenever the Bianchi identities of the fluxes hold, in the absence of sources these reduce to the following PDEs
\beq\label{eq:genclassPDES}
\frac{ \partial_y S }{K}=\text{constant}= b,~~~\partial_x^2 K+\frac{1}{v}\partial_y^2(SK)=0,~~~\partial_x^2S+\frac{b}{v}\partial_y(S K)=0,~~~v''=0.
\eeq
Notice that this is a very similar system of PDEs to that of the massive IIA class in  \eqref{eq:2.5} -  here  it is now $b$, which is  related to the NS flux through $\mathbb{T}^3$, that plays the role that $F_0$  previously did.\\
~~\\
We now need to double Wick rotate the above class to get to a class of solutions on AdS$_2\times\text{S}^3\times\mathbb{T}^3\times \Sigma_2$ - this is achieved as follows: First we wick rotate the coordinates of AdS$_2$ and S$^3$ so that they become AdS$_3$ and S$^2$ respectively. The rules are
\begin{align}
ds^2(\text{AdS}_3) & \rightarrow -ds^2(\text{S}^3), \qquad ds^2(\text{S}^2) \rightarrow -ds^2(\text{AdS}_2), \nn \\
\text{vol}(\text{AdS}_3) & \rightarrow - \text{vol}(\text{S}^3), \qquad \text{vol}(\text{S}^2) \rightarrow - i \text{vol}(\text{AdS}_2).
\end{align}
In addition to this one needs to analytically continue the various functions appearing in the class as
\begin{align}
v \rightarrow i v, \qquad K \rightarrow - i K,  \qquad S \rightarrow - i S, \qquad x \rightarrow i x,
\end{align}
so that the metric has the correct signature and is real. Finally we supplement the procedure by sending
\begin{align}
e^{-\Phi} & \rightarrow -e^{-\Phi}, \qquad F \rightarrow - F,
\end{align}
which ensures that the dilaton is real. When the dust settles we arrive at a class of AdS$_2$ solutions with the following NS sector
\begin{align}
ds^2 &= \frac{ v}{\sqrt{KS  }} \left( \frac{ 1}{4 \tilde{\Xi}} ds^2(\text{AdS}_2) + ds^2(\text{S}^3) \right) +  \sqrt{\frac{S}{K}}ds^2(\mathbb{T}^3) + \frac{ \sqrt{KS}}{v} dx^2 + \sqrt{\frac{K}{S}} dy^2, \nn \\[2mm]
e^{-\Phi} &= \frac{ K \sqrt{S} \sqrt{\tilde{\Xi}}}{\sqrt{v}},~~~ H = -\frac{1}{2} d\left(\frac{  v v'}{4 K S \tilde{\Xi}}+x\right)\wedge \text{vol}(\text{AdS}_2) +\frac{ 1 }{K} \partial_y S \text{vol}(\mathbb{T}^3),~~~\tilde{\Xi}=1-\frac{v'^2}{4 K S } \label{eq:genmetric }
\end{align}
and the following $d=10$ RR fluxes
\begin{align}
F_1 & = \partial_x K dy -\frac{1 }{v} \partial_y ( K S ) dx,~~~~F_5=  (1+\star_{10}) f_5 \nn  \\[2mm]
F_3 & = \left( \frac{ v' \partial_y ( K S )}{8 K S \tilde{\Xi}} dx -\frac{ v v' \partial_x\log K dy }{8S \tilde{\Xi}}-\frac{1}{2} K dy \right)\wedge \text{vol}(\text{AdS}_2) - \partial_x S \text{vol}(\mathbb{T}^3), \nn \\[2mm]
f_5 & = \frac{1}{ 8 K S\tilde{\Xi}} \left( 4 K S^2 + v^2 v' \partial_x \left(\frac{S}{v}\right) \right) \text{vol}(\text{AdS}_2) \wedge\text{vol}(\mathbb{T}^3) .\label{eq:genflux}
\end{align}
One has a solution to the Type IIB equations of motion whenever the Bianchi identities of the fluxes hold, these still reduce to the PDEs in \eqref{eq:genclassPDES} away from the loci of sources. This class of AdS$_2$ solutions should preserve small ${\cal N}=4$ supersymmetry like the AdS$_3$ class we generated it from.

In the next sections we show how to recover the solutions of section \ref{eq:case1SL2Tdual} and \ref{eq:case2SL2Tdual} from this more general class of AdS$_2$ solutions. This actually also serves as evidence to our claim that the general class preserves small ${\cal N}=4$ supersymmetry, because these two specific solutions certainly do.

\subsubsection{Recovering the $h g\propto 1$ case}
The solution of section  \ref{eq:case1SL2Tdual} is recovered from \eqref{eq:genmetric }-\eqref{eq:genflux} by fixing
\begin{align}
S = c q v_0(x), ~~~~ K = \frac{q}{c} p(y) v_0(x),~~~~ v=q^2v_0,
\end{align}
where the functions $v_0,p$ obey
\beq
\partial_x^2v_0=0,~~~~ \partial_{y}^2 p=0,
\eeq
which solves all of \eqref{eq:genclassPDES} for $b=0$. To precisely reproduce the form of the solution in  \ref{eq:case1SL2Tdual} one needs to do an implicit coordinate transformation such that
\beq
p(y)= \sqrt{u(\rho)}.
\eeq
The resulting solution is actually a mild generalisation of the solution in section  \ref{eq:case1SL2Tdual} which depends on the linear function $v_0$. Specifically if one fixes $v_0=x$ one recovers the solution of section  \ref{eq:case1SL2Tdual}, but if one fixes $v_0=$ constant one instead recovers its U(1)  T-dual analogue. Such hybrid T-dual/non-Abelian T-dual solutions were previously found in  \cite{Lozano:2019emq}.

\subsubsection{Recovering the $h=$ constant case}
Likewise the solution of section  \ref{eq:case2SL2Tdual} is recovered from \eqref{eq:genmetric }-\eqref{eq:genflux} by fixing
\begin{align}
S=  q \sqrt{Q(y)} v_0(x),~~~~K= \frac{h_0 q v_0(x)}{\sqrt{Q(y)}},~~~~v=q^2v_0(x),
\end{align}
where 
\beq
\partial_{x}^2v_0=0,~~~ \partial_y Q= 2bh_0,
\eeq
which solves \eqref{eq:genclassPDES}. To get to the precise form of the solution in section \ref{eq:case2SL2Tdual}
one needs to change coordinates from $y$ to $\rho$ in such a way that
\begin{align}
Q(y)= h_0^2 g(\rho)^2
\end{align}
which requires that one identifies
\beq
b= h_0 \partial_{\rho}g.
\eeq
Again the result of doing this is actually a T-dual/non-Abelian T-dual hybrid of the solution we seek: Fixing $v_0= x$ recovers the solution of section  \ref{eq:case2SL2Tdual} while $v_0=$ constant gives rise to its U(1) T-dual analogue.

\section{Field theory analysis}\label{field-theory}

In this section we study the 1d SCFTs dual to the $\text{AdS}_2\times \text{S}^3\times \mathbb{T}^3\times \Sigma_2$ solutions with $F_0=0$ constructed in section \ref{eq:case2SL2Tdual}, as a first step towards a more general analysis of the field theories dual to our broader class of $\mathcal{N}=4$ solutions, in particular the ones constructed in section \ref{eq:case1SL2Tdual}. We construct explicit quivers that we conjecture flow in the IR to the SCQMs dual  to the solutions, and check our proposal with the computation of the field theory and holographic central charges. We propose an interpretation of the solutions as holographic duals of monopole bubbling in 4d $\mathcal{N}=2$ theories living in D3-D7 branes, and of baryon vertices in 5d $\mathcal{N}=1$ theories living in D5-NS5-D7 branes. These two interpretations are based on the disposition of the branes that make the brane set-up associated to the solutions along the two non-compact directions of the 2d Riemann surface.  

\subsection{Hanany-Witten brane set-up and quantised charges}\label{charges}

Even if we have not constructed the brane solution whose near horizon geometry is described by the geometry specified  by equations \eqref{eq:4.1} and \eqref{eq:4.1b}\footnote{This is a difficult task when a non-Abelian T-duality transformation is involved (see \cite{Terrisse:2018hhf,Dibitetto:2020bsh}).}, one can see that the brane intersection shown in Table \ref{D1D7NS5D5D3F1}
is consistent with both the fluxes and the preserved supersymmetries of the solutions. As it typically happens the $x^i$ coordinates will mix upon taking the near horizon limit to give rise to the radius of AdS, the two non-compact directions $r$ and $\rho$ and the coordinates along the $\text{S}^3$. 
In this configuration 
the D1-branes play the role of colour branes, and $x^4$  of field theory direction, and the F1-strings stretch along $x^5$. As we will show below once the near horizon limit is taken the D1-branes will effectively stretch along the $\rho$ direction and the F1-strings along $r$.
We will also see that the D1-branes carry as well electric charge, which allows one to interpret them as baryon vertices for the D7-branes. The same interpretation is found for the D3-branes with respect to the D5-branes. We can indeed check in Table \ref{D1D7NS5D5D3F1} that these branes have the right relative orientations that allow to infer this. Alternatively, as we will discuss, the brane set-up can be interpreted as 
describing the super conformal quantum mechanics associated to the bubbling of dyonic monopoles in the 4d theory living in the intersection of the D3 and D7 branes, in a brane scenario in which the D3-D7 branes and the D1-F1 strings are put inside a 5-brane web. 
 \begin{table}[http!]
	\begin{center}
		\begin{tabular}{| l | c | c | c| c | c | c | c | c| c | c |}
			\hline		    
			& $t$ & $z_1$ & $z_2$ & $z_3$ & $x^4$ & $x^5$ & $x^6$ & $x^7$ & $x^8$ & $x^9$ \\ \hline
			D1 & x &   & & & x & & & & &\\ \hline
			D3 & x & x & x & x & & & & & &\\ \hline
			D5 & x &   &  &  &x & & x  &x   &x   &x   \\ \hline
			D7 & x &  x & x & x & & &x &x &x &x \\ \hline
			NS5 & x &  &   &   & & x & x  & x  & x & x  \\ \hline
			F1 & x & &  &    & & x &  &  & &  \\ \hline
		\end{tabular} 
	\end{center}
	\caption{$\frac18$-BPS brane intersection underlying the $\mathcal{N}=4$ AdS$_2$ solutions constructed in section \ref{eq:case2SL2Tdual}. $t$ is the time direction where the 1d dual CFT lives; ($z_1$, $z_2$, $z_3$) parametrise the $\mathbb{T}^3$; $x^4$ is the field theory direction; $x^5$ is the direction where the F1-strings lie, and ($x^6,x^7,x^8,x^9$) are the directions associated to the isometries of the S$^3$.}
		\label{D1D7NS5D5D3F1}	
\end{table}


Let us start our discussion with the computation of the quantised charges. To this end we find it helpful to decompose the NS 3-form into its electric and magnetic parts as
\beq\label{eq:H3decompose}
H=H^e+ H^m,~~~~ H^e=-\frac{1}{2}d\left(\frac{r}{1-  \frac{q^2}{4 h_0^2r^2}}\right)\wedge \text{vol}(\text{AdS}_2),~~~~H^m=h_0 g' \text{vol}(\mathbb{T}^3).
\eeq
We start analysing the D7-D5-NS5 brane subsystem. The quantised charge associated to the D7-branes is obtained integrating the magnetic component of $F_1$, as
\begin{equation}\label{D7charge}
Q_{D7}=\int f_1=q\int d\rho.
\end{equation}
To proceed we need to define the domain of integration of the $\rho$-direction, which is in turn related to the way we define the Page fluxes in our backgrounds. Indeed, the standard definition of Page fluxes given by 
\eqref{eq:standardPagecharges} cannot be used for our solutions, since there is no globally defined magnetic $B$-field.
Instead, we note that the definition\footnote{We use $f_p$ to denote $p$-form RR magnetic fluxes, and an $m$ superscript to denote magnetic components.}
\begin{equation}\label{newPage}
\hat{f}_p=f_p+H^m\wedge C^m_{p-3},
\end{equation}
is equivalent to  \eqref{eq:standardPagecharges} for the magnetic part of the Page fluxes. This definition of the Page fluxes has the non-trivial implication that instead of having Dp-branes being created as NS5-branes are crossed, which is the usual way the Hanany-Witten brane creation effect is understood at the level of the fluxes, it describes Dp-branes being created as D(p+2)-branes are crossed. In particular, for the D5-branes we have
\begin{equation}\label{newPageD5}
\hat{f}_3=f_3+H^m\,C_{0}.
\end{equation}
From this expression we see that $\hat{f}_3$ is sensitive to gauge transformations of $C_0$. In order to carefully account for these we demand that $C_0$ lies in the range $C_0\in [0,q]$. In order to accomplish this we need to take
\begin{equation}\label{C0mod}
C_0=q(\rho-k) \qquad \text{for} \qquad \rho\in [k,k+1].
\end{equation}
Then, given that the D7-brane charge is obtained computing
\begin{equation}
Q_{D7}^{(k)}=\int_{k}^{k+1} dC_0,
\end{equation}
this imposes that $q$ D7-branes are created in this interval. This clarifies the role played by the large gauge transformations of $C_0$, as generating a strong coupling realisation of the Hanany-Witten brane creation effect by which D7-branes, instead of NS5-branes, are positioned along the field theory direction. In turn, the large gauge transformation in \eqref{C0mod} modifies $\hat{f}_3$ as $\hat{f}_3\rightarrow \hat{f}_3-q kH^m$ such that
\begin{equation}
\hat{f}_3=qh_0 \Bigl(g' (\rho-k)-g\Bigr)\text{vol}(\mathbb{T}^3) \qquad \text{in} \qquad \rho\in [k,k+1].
\end{equation}
\noindent Recall that away from the loci of sources the Bianchi identities require that $g''=0$, which makes $g$ a linear function. However globally we need only  impose that $g$ is continuous, while $g'$ is allowed to have discontinuities that give rise to delta function sources in $g''$, signaling the presence of smeared NS5-branes at $\rho_k=k$. Therefore, we take $g$ piecewise linear such that
\begin{equation}
g_k=\frac{1}{2\pi h_0}\Bigl(\alpha_k+\beta_k (\rho-k)\Bigr) \qquad \text{for} \qquad \rho\in [k,k+1].
\end{equation}
Here $\alpha_k$ and $\beta_k$ are integer numbers, since they are related to the numbers of D5 and NS5-branes in the $k$th interval, according to
\begin{eqnarray}
Q_{D5}^{(k)}&=&-\frac{1}{(2\pi)^2}\int_{\mathbb{T}^3} \hat{f}_3=q \alpha_k,\label{QD5k}\\
Q_{NS5}^{(k)}&=&\frac{1}{(2\pi)^2}\int_{\mathbb{T}^3}H^m=\beta_k.
\end{eqnarray}
As well defined global solutions should have a continuous metric and dilaton this amounts to requiring that $\alpha_k, \beta_k$ must satisfy
\begin{equation}
\alpha_{k+1}=\alpha_k+\beta_k=\sum_{j=0}^{k}\beta_j.
\end{equation}
The value of $\beta_k$ is not fixed however, indeed we allow it to change between intervals, this means that in the vicinity of $\rho=k$ we have
\beq
g''= \frac{1}{2\pi h_0} (\beta_k-\beta_{k-1})\delta(\rho-k)
\eeq
and so we have source NS5 branes with charge $\beta_k-\beta_{k-1}$ at  $\rho=k$ that are extended in AdS$_3\times\text{S}^3$ and smeared over $\mathbb{T}^3$. Note that this does not induce any source terms in the RR sector as should be clear from examining \eqref{eq:4.1b}, which is independent of $g'$.
Next we need to care about how the space ends on the $\rho$-direction. We choose to take $\rho\in [0,P+1]$ and impose that at both ends $g$ vanishes. This implies that in the $[0,1]$ and $[P,P+1]$ intervals we must have
\begin{equation}
g_0=\frac{\beta_0}{2\pi h_0}\rho, \qquad
g_P=\frac{1}{2\pi h_0}\Bigl(\alpha_P+\beta_P (\rho-P)\Bigr),
\end{equation}
with $\beta_P=-\alpha_P$.
The behaviour close to both ends of the space is then that of an ONS5 orientifold fixed plane (S-dual of an O5) that is smeared on the $\mathbb{T}^3$. 

\noindent Summarising our results so far, we have seen that there are $q \alpha_k$ D5-branes stretched between $q$ D7-branes located at $\rho_k=k$, $\rho_{k+1}=k+1$ with $\beta_k$ perpendicular NS5-branes in each $[\rho_k,\rho_{k+1}]$ interval. 

Let us analyse now the D1-D3-F1 subsector of the brane set-up. As we have mentioned, the D1 and D3-branes carry electric charges, which allows us to interpret them as baryon vertices. 
These charges are obtained integrating the Page fluxes associated to the electric components of ${F}_3$ and ${F}_5$ in \eqref{eq:4.1b}, defining these is a little subtle in this case. The electric part of $H$ in \eqref{eq:H3decompose} allows for a globally defined NS 2-form, given by 
\begin{equation}\label{B2electric}
B^e=-\frac{1}{2}\frac{r}{1- \frac{q^2}{4h_0 r^2}}\, \text{vol}(\text{AdS}_2).
\end{equation}
We need to define some combinations of the fluxes whose integrals maybe be interpreted as charges, as such they should be closed (away from sources). It is not hard to see that if one decomposes the magnetic fluxes as\footnote{We use  an $e$ superscript to denote electric components.}
\beq
F^e_p=  dC^e_{p-1}- H^m\wedge  C^e_{p-3}+B^e\wedge f_{p-2}
\eeq
for $C^e_p$ some electric RR potentials, we get the correct Bianchi identity for the electric part of the flux, ie  $dF^e_p=  H^m\wedge F^e_{p-2}+H^e\wedge f_{p-2}$. This suggests defining the electric Page fluxes as
\beq\label{electricPagefluxes}
\hat{F}_p^e=F_p^e+H^m\wedge  C^e_{p-3}-B^e\wedge f_{p-2},\\
\eeq
which are closed (away from sources) by construction.
In turn, the F1-strings are electrically charged with respect to $B$, with charge 
\begin{equation}
Q_{F1}^{e}=\frac{1}{(2\pi)^2}\int_{\text{AdS}_2}B^e,
\end{equation}
with $B^e$ given by \eqref{B2electric}. One can easily see that regularising the volume of $\text{AdS}_2$ as $\text{Vol}(\text{AdS}_2)=4\pi$ and dividing the $r$ direction in intervals of length $2\pi$ a F1-string lies at each such interval. This can be linked to the condition that the integral of $B^e$ lies in the fundamental region. In this case a large gauge transformation of gauge parameter $n$ must be performed for $r\in [2n\pi,2(n+1)\pi]$, such that
\begin{equation}\label{B2electriclarge}
B^e=\left(-\frac{1}{2}\frac{r}{1- \frac{q^2}{4h_0 r^2}}+n\pi\right) \text{vol}(\text{AdS}_2),
\end{equation}
and an F1-string is created. We identify the beginning of the space at $r_0=\frac{q}{2\sqrt{h_0}}$ with $2\pi$, consistently with the fact that, as we showed in section \ref{generalsols}, the behaviour of the metric and dilaton at this point is that associated to an OF1-plane. This fixes
$h_0=\frac{q^2}{16\pi^2}$.

\begin{figure}
\centering
\includegraphics[scale=0.4]{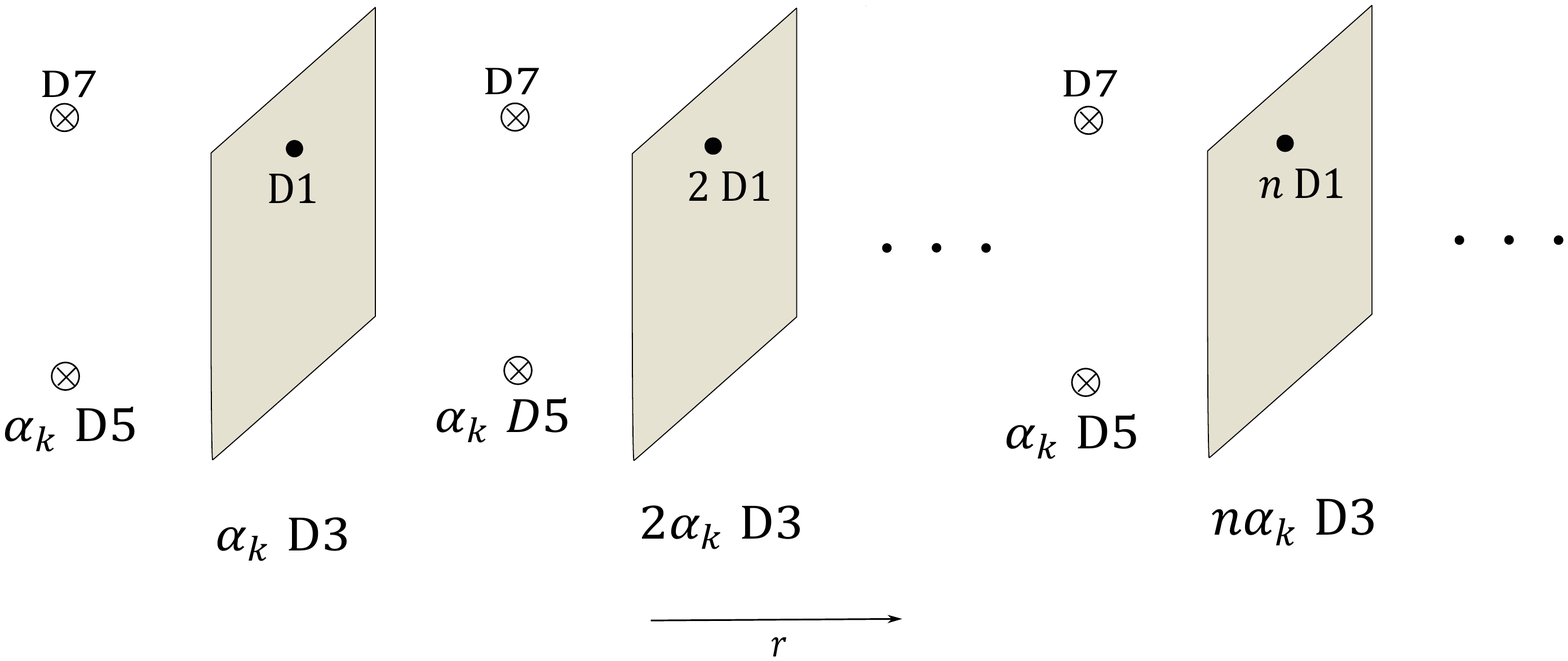}
\vspace{-2cm}
\caption{Brane set-up in the $r$ direction for $\rho$ constant, in units of $q$. The numbers of D7 and D5 branes at each interval are given by their respective magnetic charges. Instead, for the numbers of D1 and D3 branes we give their electric charges (computed in \eqref{QD1e} and \eqref{QD3e}) as these are the ones that play a role in their interpretation as baryon vertices.}
\label{brane-set-up-r}
\end{figure}

\noindent Substituting in \eqref{electricPagefluxes} we then find in the interval $r\in [2n\pi,2(n+1)\pi]$ that
\begin{eqnarray}
\hat{F}_3^e&=&- n\pi q\, d\rho\wedge \text{vol}(\text{AdS}_2),~~~\Rightarrow  ~~~C_2^e=  -2\pi n  q (\rho-l)\nn\\
\hat{F}_5^e&=& \frac{1}{2} n q (g+(\rho- l) g')\text{vol}(\text{AdS}_2)\wedge \text{vol}(\mathbb{T}^3)
\end{eqnarray}
where $l$ is an integration constant. Given this we find
\beq
Q_{D1}^{e(n)}=-\frac{1}{(2\pi)^2}\int \hat{F}_3^e=nq=n Q_{D7}^{(k)}\label{QD1e}, 
\eeq
meaning that there are $nq$ units of electric D1 charge within the cell $r\in [2n\pi,2(n+1)\pi]$, thus for consistency we should  constrain
\beq
0<-C_2^e< n q~~~~\Rightarrow~~~~ l=k.
\eeq
We then have that the electric D3 brane charge is given by
\beq
Q_{D3}^{e(n)}=\frac{1}{(2\pi)^4}\int  \hat{F}_5^e=nq \alpha_k = n Q_{D5}^{(k)}\label{QD3e}.
\eeq
 The brane set-up associated to these quantised charges is depicted in Figure \ref{brane-set-up-r}.

\begin{figure}
\centering
\includegraphics[scale=0.65]{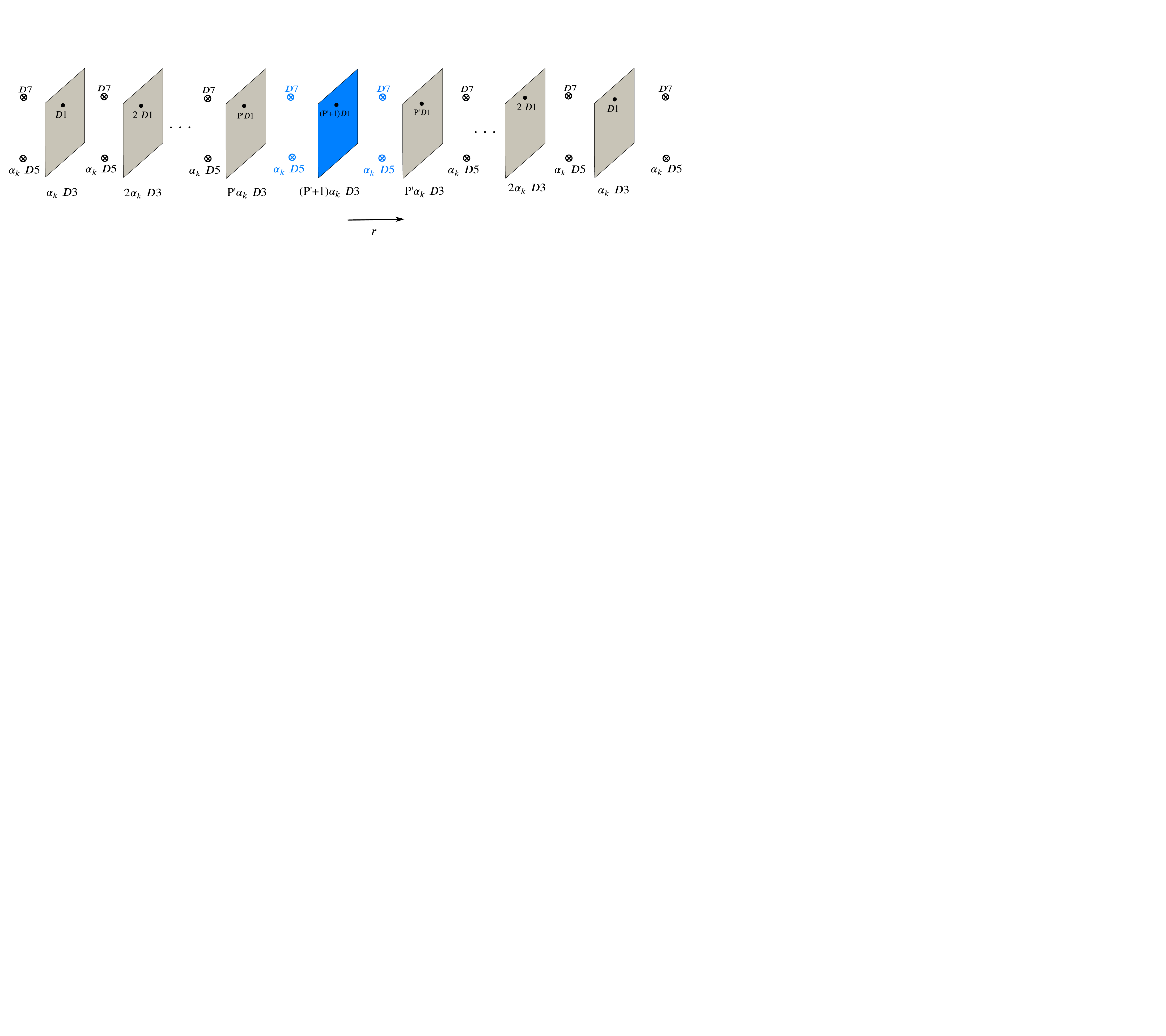}
\vspace{-21cm}
\caption{Completed brane set-up in the $r$ direction for $\rho$ constant, in units of $q$. The branes in blue are used to glue the brane set-up with itself.}
\label{completed-brane-set-up}
\end{figure}

\noindent Clearly, as $r$ is a non-compact direction we need to care about its global definition. This will be important later on when we compute the central charge associated to the solutions. We choose to complete the solutions in this direction by glueing them to themselves, at a certain point $r_{P'+1}= 2(P'+1)\pi$. The completed brane set-up is depicted in Figure \ref{completed-brane-set-up}. 
Specifically, the solutions are extended beyond the $[2\pi,2(P'+1)\pi]$ interval by gluing a second solution onto it which takes the form of \eqref
{eq:4.1}-\eqref{eq:4.1b}, but with  $r\rightarrow 4(P'+1)\pi-r$ and for which  $r\in [2(P'+1)\pi,2(2P'+1)\pi]$. One can easily check that the metric and fluxes of this global completion are continuous across $r=2(P'+1)\pi$. The $r\in [2(P'+1)\pi,2(2P'+1)\pi]$ part of the interval  is likewise divided in $P'$ intervals of length $2\pi$ where F1-strings are created, associated to the large gauge transformations of the $B^e$ field, that we complete as
 \begin{equation}
B^e =  \left\{ \begin{array}{ccrcl}
\Bigl(-\frac{1}{2}\frac{r^3}{r^2-(2\pi)^2}+n\pi\Bigr)\text{vol}(\text{AdS}_2)\,,   \quad r\in [2\pi, 2(P'+1)\pi]\\[2mm]
\Bigl(-\frac{1}{2}\frac{(4(P'+1)\pi-r)^3}{(4(P'+1)\pi-r)^2-(2\pi)^2}+(4(P'+1)-n)\pi\Bigr)\text{vol}(\text{AdS}_2)\,, \quad r\in [2(P'+1)\pi, 2(2P'+1)\pi]
\end{array} 
\right.
\end{equation}
Note that with this completion of the solutions a new singularity associated to an OF1-plane arises at $r=2(2P'+1)\pi$, where the space now ends.

\subsection{Baryon vertex interpretation}\label{baryon-vertex-interpretation}

\begin{figure}
\centering
\includegraphics[scale=0.45]{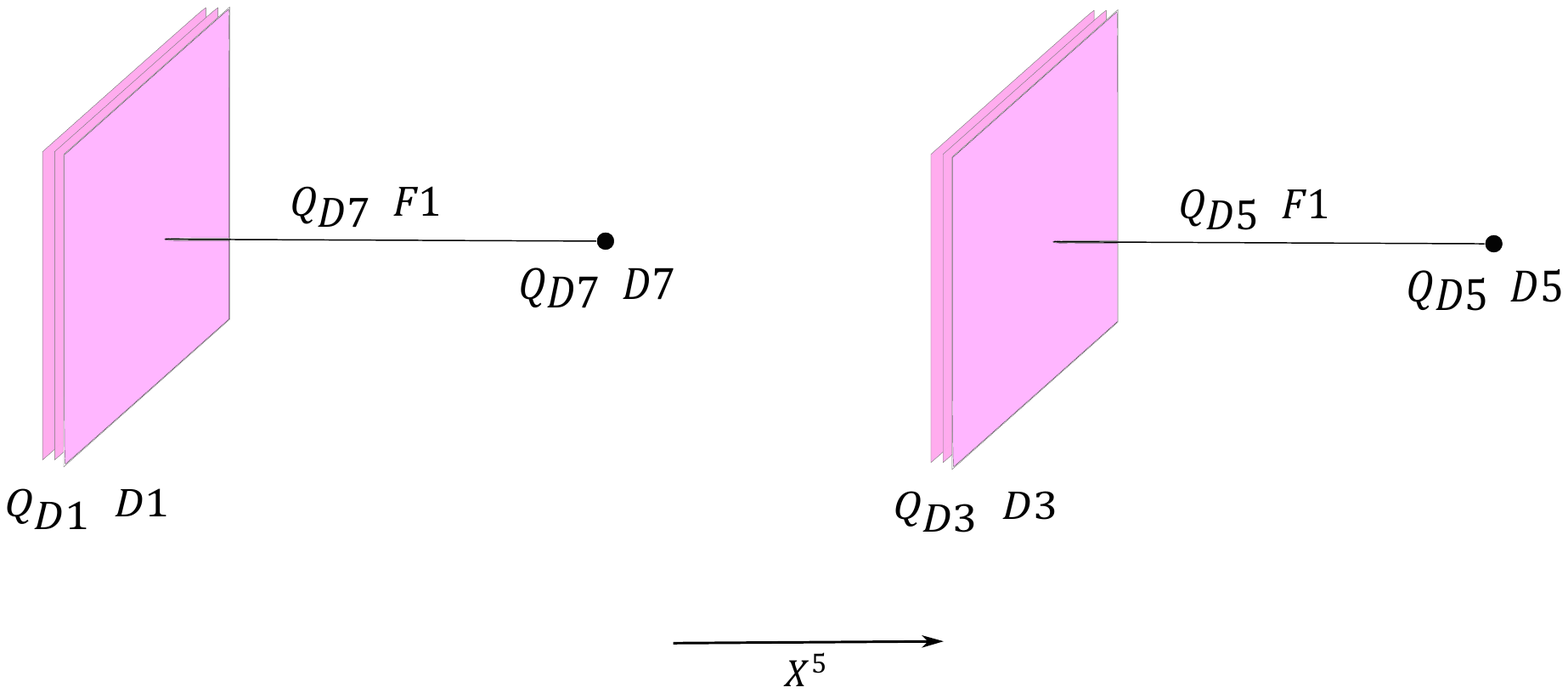}
\vspace{-8cm}
\caption{Wilson loops in the $Q_{D7}$-th and $Q_{D5}$-th antisymmetric representations of U($Q_{D1}$) and U($Q_{D3}$), respectively.}
\label{baryon-vertex}
\end{figure}

Our previous analysis shows that there is one F1-string extended in $r$ for $r\in [2n\pi,2(n+1)\pi]$ for $n=1,2,\dots 2P'$. At each of these intervals there are $nq$ units of D1-brane electric charge and $nq\alpha_k$ units of D3-brane electric charge for $n=1,2,\dots, P'+1$, as well as $(4(P'+1)-n)q$ units of D1-brane electric charge and $(4(P'+1)-n)q\alpha_k$ units of D3-brane electric charge for $n=P'+1,P'+2,\dots, 2P'+1$.
Recalling the existence of the following couplings in the WZ actions of the D1 and D3 branes,
\begin{equation}
S_{D1}=T_1\int f_1\wedge A_t, \qquad S_{D3}=T_3\int \hat{f}_3\wedge A_t,
\end{equation}
where $A_t$ is the electric component of the Born-Infeld vector field, we find that a D1-brane extended in $\rho$ between $[\rho_k,\rho_{k+1}]$ and located at a fixed position in the $r\in [2n\pi,2(n+1)\pi]$ interval behaves as a baryon vertex for the D7-branes, since it carries $Q_{D1}^e=nq=nQ_{D7}$ units of F1-string charge if $n=1,2,\dots, P'+1$, and $Q_{D1}^e=(4(P'+1)-n)Q_{D7}$ if $n=P'+1,\dots, 2P'+1$. Similarly, a D3-brane wrapped on the $\mathbb{T}^3$ and located at a fixed position in the $r\in [2n\pi,2(n+1)\pi]$ interval carries $Q_{D3}^e=n q\alpha_k=nQ_{D5}$ units of F1-string charge if $n=1,2,\dots, P'+1$ and $Q_{D3}^e=(4(P'+1)-n)Q_{D5}$ if $n=P'+1,\dots, 2P'+1$, and therefore behaves as a baryon vertex for the D5-branes. Indeed, the relative orientation between the D1 and the D7 branes and between the D3 and the D5 branes in the brane set-up is the one that allows to create F1-strings stretched between the D1 and the D7 branes and between the D3 and the D5 branes, as depicted in Figure \ref{baryon-vertex}. 

\begin{figure}
\centering
\includegraphics[scale=0.55]{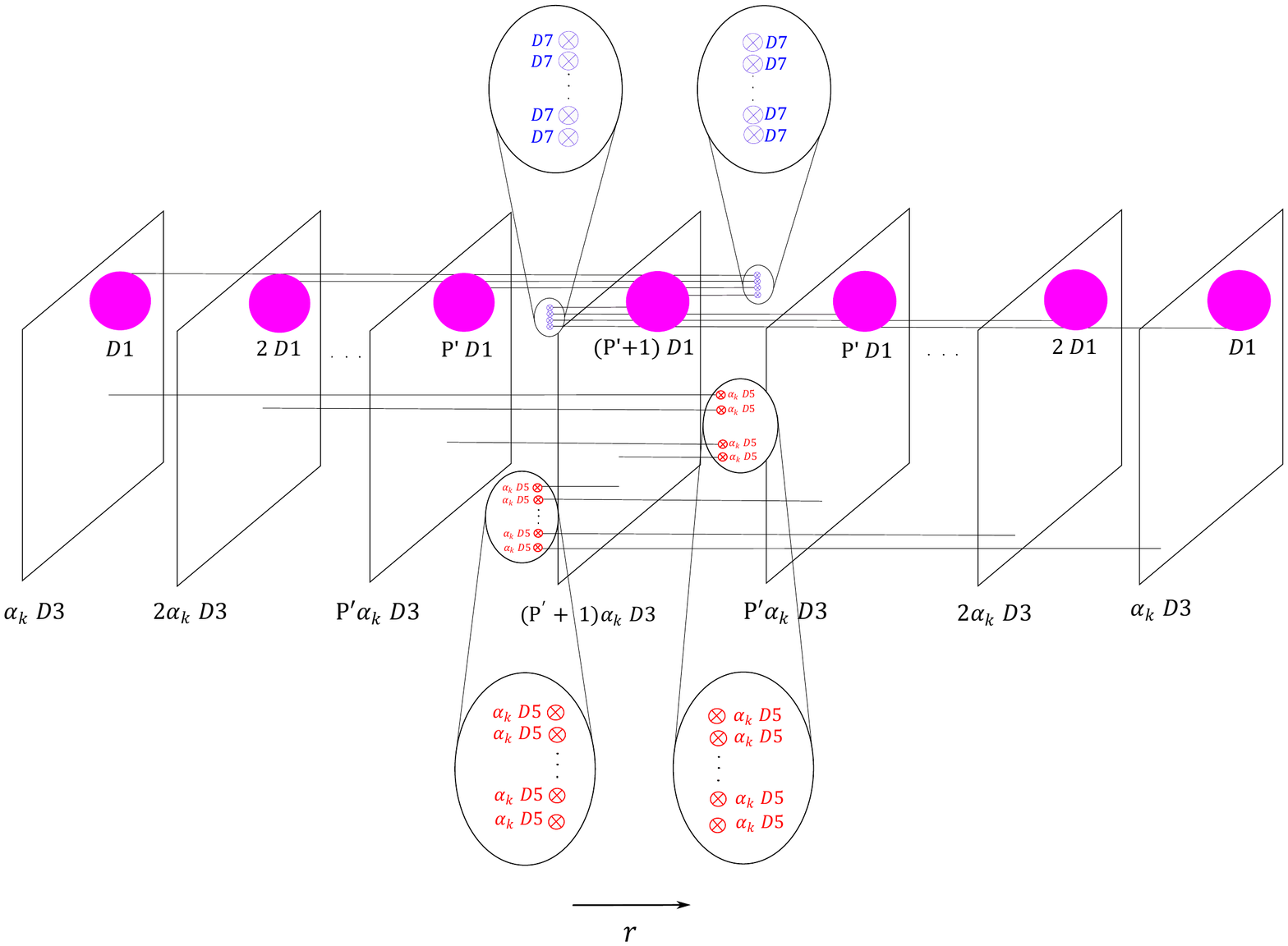}
\vspace{-2.25cm}
\caption{Hanany-Witten brane set-up equivalent to the brane configuration in Figure \ref{completed-brane-set-up}, in units of $q$.}
\label{HW-moves}
\end{figure}

\begin{figure}[H]
\centering
\includegraphics[scale=0.5]{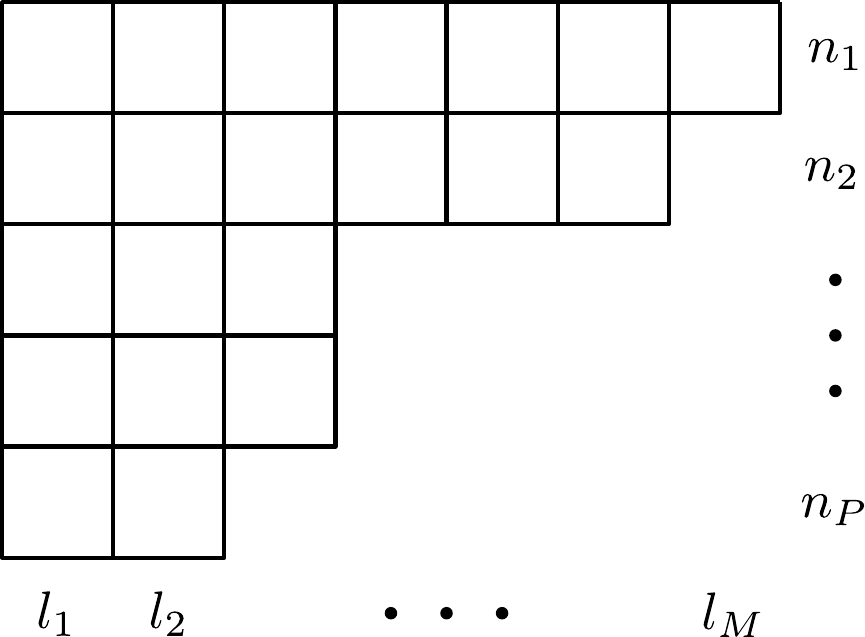}
\caption{Young tableau labelling the irreducible representations of U(N).}\label{youngtableau}
\end{figure}

As we have mentioned, the location of the branes along the $r$ direction is the one depicted in Figure \ref{completed-brane-set-up}. As explained in detail in \cite{Lozano:2020sae} this brane configuration can be related by a combination of a T-duality, an S-duality, successive Hanany-Witten moves and a further T-duality to the brane set-up depicted in Figure \ref{HW-moves}.
After these transformations the analogy with the description of half-BPS Wilson loops in antisymmetric representations labelled by the Young tableau depicted in Figure \ref{youngtableau}, proposed in \cite{Yamaguchi:2006tq,Gomis:2006sb}, is evident. 
\noindent In these references if was shown that a Wilson loop in the $(l_1,l_2,\dots,l_M)$ antisymmetric representation is realised by a configuration of stacks of branes separated a distance $L$ from the colour branes with $(l_1,l_2,\dots,l_M)$ F1-strings stretched between the stacks.
This is completely analogous to our configuration in Figure \ref{HW-moves}. In our case we encounter the particular situation in which the sum of the F1-strings stretched between each D1 (D3) and the flavour D7 (D5) branes coincides with the rank of the gauge group of the D1 (D3) branes, which implies that the Wilson lines are in the fundamental representation of the gauge groups. Therefore, the D1-D3 branes behave as baryon vertices for the D7-D5 branes. 


Our analysis in this subsection suggests that the AdS$_2$ solutions constructed in section \ref{eq:case2SL2Tdual} could find an interpretation as backreacted D1-D3-F1 baryon vertices in the 5d $\mathcal{N}=1$ theory living in a D5-D7-NS5 brane intersection. Indeed, our discussion so far mimics the interpretation of the AdS$_2$ solutions found in \cite{Lozano:2020sae,Lozano:2021rmk,Ramirez:2021tkd,Lozano:2021fkk,Lozano:2022vsv}, as dual to baryon vertices in 5d Sp(N), 4d $\mathcal{N}=4$ and 4d $\mathcal{N}=2$ superconformal field theories. However, a key difference with our construction in this paper is that the AdS$_2$ solutions discussed in these references asymptote locally to the $\text{AdS}_6$ or $\text{AdS}_5$ solutions dual to the higher dimensional SCFTs where the baryon vertices are embedded, while this is not the case for our AdS$_2$ solutions. Based on this it is likely that in our construction in this paper the branes could play a different role, a possibility that we will further discuss in section \ref{thooft}, where we will see that they allow for an alternative interpretation  in terms of dyonic monopoles in the 4d theory living in D3-D7 branes. 

\subsection{Quiver quantum mechanics}\label{quiver}

In the previous subsection we computed the electric charges of the D1 and D3 branes, which are the relevant ones for the construction of Wilson lines or baryon vertices. In this subsection we are interested instead in the construction of the quiver quantum mechanics dual to our solutions, for which the relevant charges are the magnetic ones. 

Analogously to our discussion for the D7-D5-NS5 brane subsystem, the D1 and D3-branes are magnetically charged with respect to the Page fluxes defined by
\begin{equation}
\hat{f}_7=f_7+H^m\wedge C_4^m, \qquad \hat{f}_5=f_5+H^m\wedge C_2^m.
\end{equation}
These expressions show that the numbers of D1 and D3 branes are sensible to gauge transformations of $C_4$ and $C_2$, respectively. In particular, we have for our background
\begin{equation}
\hat{f}_5=dC_4^m=2qr \text{vol}(\text{S}^3)\wedge d r\wedge d\rho
\end{equation}
and, upon integration
\begin{equation}
C_4^m=2qr\rho  \text{vol}(\text{S}^3)\wedge dr.
\end{equation}
Taking $\rho\in [k,(k+1)]$ a large gauge transformation of parameter $k$, such that
 \begin{equation}
C_4^m=2qr(\rho-k)\wedge \text{vol}(\text{S}^3)\wedge  dr,
\end{equation}
creates a number of D3-branes in this interval given by
\begin{equation}\label{QD3intervals}
Q_{D3}^{(k)}= q \left\{ \begin{array}{ccrcl}
n+\frac12, \quad  \text{for} \quad n=1,2,\dots, P'\\
4(P'+1)-n-\frac12, \quad \text{for} \quad n=P'+1,\dots, 2P'\, .
\end{array}
\right.
\end{equation}
Similarly, substituting in 
$\hat{f}_7$ above we find,
\begin{equation}
\hat{f}_7=2q h_0 r\Bigl(g-g'(\rho-k)\Bigr)dr\wedge \text{vol}(\mathbb{T}^3)\wedge \text{vol}(\text{S}^3),
\end{equation}
from which we compute
\begin{equation}\label{QD1intervals}
Q_{D1}^{(k)}=\frac{1}{(2\pi)^6}\int \hat{f}_7= q \alpha_k . \left\{ \begin{array}{ccrcl}
n+\frac12, \quad  \text{for} \quad n=1,2,\dots, P'\\
4(P'+1)-n-\frac12, \quad \text{for} \quad n=P'+1,\dots, 2P'\, .
\end{array}
\right.
\end{equation}




\noindent Combining with our results for the magnetic charges for the D7 and D5 branes, reported in subsection \ref{charges}, the picture that arises is a non-perturbative setting in which $q \alpha_k$ D5-branes are stretched between stacks of $q$ D7-branes located at $\rho_k=k$ and $\rho_{k+1}=(k+1)$, and $q (n+\frac12) \alpha_k\equiv qp\alpha_k$ D1-branes are stretched between stacks of $q p$ D3-branes located at the same positions. This is so for $r\in [2n\pi, 2(n+1)\pi]$ intervals with $n=1,2\dots, P'$, with the obvious changes implied by equations \eqref{QD3intervals} and \eqref{QD1intervals} in the intervals with  $n=P'+1,\dots, 2P'$. 

In this non-perturbative brane scenario D1-branes are created when D3-branes are crossed, while D5-branes are created when D7-branes are crossed. These are non-perturbative realisations of Hanany-Witten brane set-ups, in which Dp-branes are stretched between D(p+2)-branes located at fixed positions in the field theory direction with orthogonal NS5-branes lying between the stacks of D(p+2)-branes. In our case the D5-branes will finally contribute with flavour groups, leaving just the D1-branes stretched between D3-branes as the ones giving rise to colour groups. Note that the configuration consisting on D1-branes stretched between D3-branes is precisely the Type IIB description of smooth monopoles in the 4d theory living in D3 or D3-D7 brane systems, according to the Nahm construction \cite{Diaconescu:1996rk}. Our solutions describe naturally this type of configurations, because of the definition of Page fluxes that needs to be taken in the absence of a globally well-defined $B$ field.  We will come back to this discussion in subsection \ref{thooft} when we relate our SCQM to the bubbling of singular loops in the 4d theory living in D3-D7 branes.

From the previous quantised (magnetic) charges we can now proceed with the construction of the quiver that describes the supersymmetric quantum mechanics that, we propose, flows in the IR to the SCQM dual to our solutions. In order to extract it we need to account for the ordering of the NS5-branes along the $\rho$-direction, together with the net number of D1-branes ending on them and the number of orthogonal D3-branes lying between NS5-branes. Similarly, in order to account for the number of D5-branes in each interval we need to compute the net number of them ending on NS5-branes together with the number of orthogonal D7-branes between NS5-branes. The massless modes that give rise to the quiver quantum mechanics arise then from the open strings that connect the D-branes in the same interval between NS5-branes or adjacent ones. 

\begin{figure}
\centering
\includegraphics[scale=0.55]{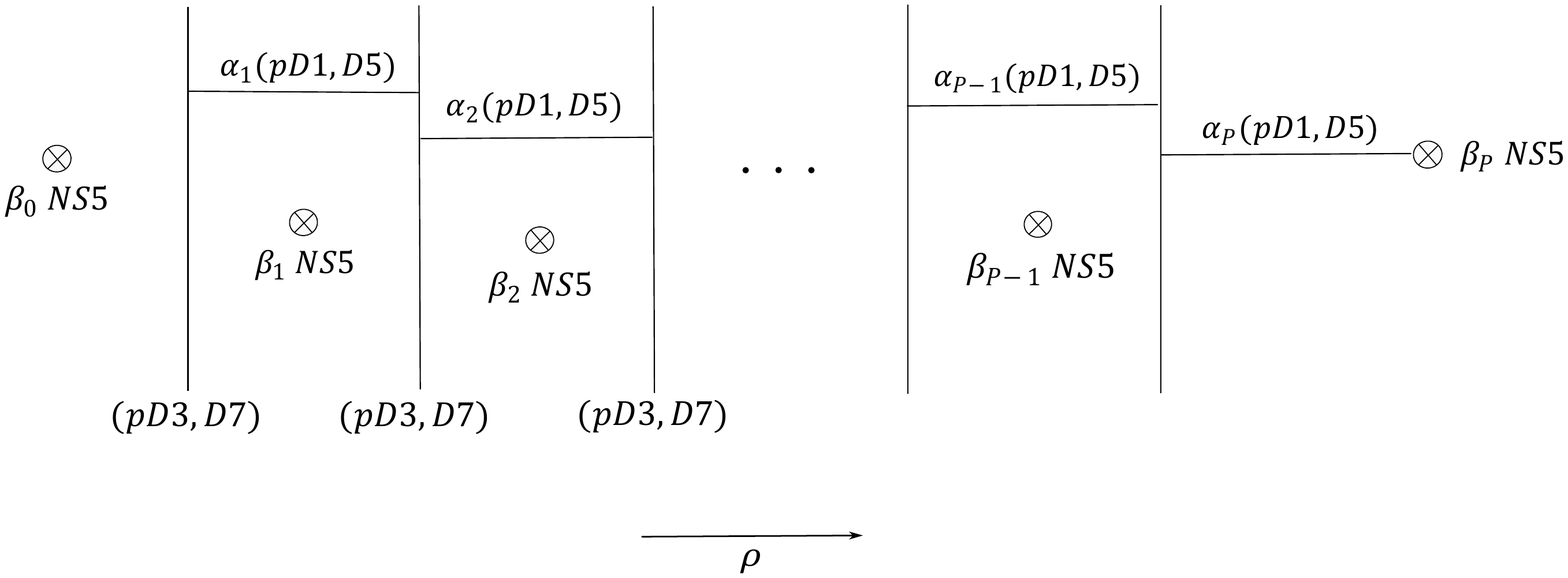}
\vspace{-5cm}
\caption{Brane set-up along the $\rho$ direction, for $r$ constant, in units of $q$.}
\label{brane-set-up-rho}
\end{figure}

Given that in our brane set-up, depicted in Figure \ref{brane-set-up-rho}, the D1 and D5 branes stretch, respectively, between D3 and D7 branes with NS5-branes orthogonal to them, we need to perform a series of Hanany-Witten moves that create D1 and D5 branes  stretching between NS5-branes, with orthogonal D3 and D7 branes lying between them. 
This is explained in detail in Appendix \ref{fieldtheory}, following closely \cite{Lozano:2022ouq}, where the same analysis was carried out for the D2-D4-NS5 brane system 
underlying the $\text{AdS}_3$ solutions studied therein. We will find the same field content as in the 2d D2-D4-NS5 brane system discussed in \cite{Lozano:2022ouq} with the D2-branes replaced by D1 (or wrapped D5) branes and the D4-branes by D3 (or wrapped D7) branes, with 1d $\mathcal{N}=4$ multiplets described in terms of 2d (0,4) multiplets in the standard way.

\begin{figure}
\centering
\includegraphics[scale=0.65]{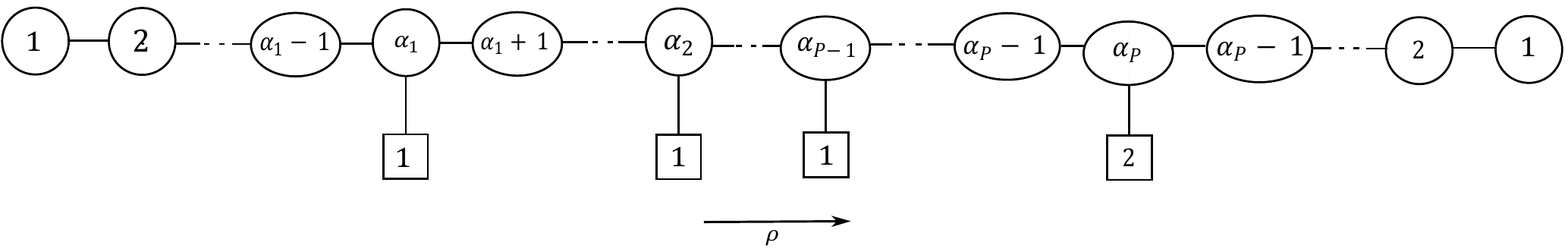}
\vspace{-17cm}
\caption{Quiver associated to the D5-D7-NS5 brane subsystem for $r$ constant. Circles denote (4,4) vector multiplets and black lines (4,4) bifundamental hypermultiplets. The gauge groups with ranks $\alpha_k$, with $k=1,\dots, P-1$ to the left of the gauge group with rank $\alpha_P$ have U(1) flavour symmetries. The gauge group with rank $\alpha_P$ has U(2) flavour symmetry. The rest of gauge groups do not have attached any flavours. We have used units of $q$.}
\label{quiverD5D7}
\end{figure}

\begin{figure}
\centering
\includegraphics[scale=0.65]{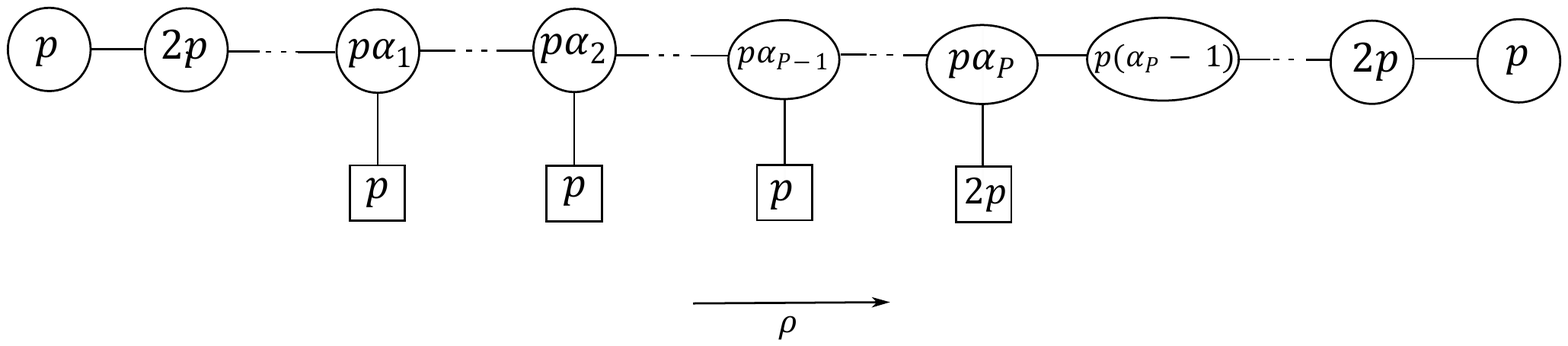}
\vspace{-16.25cm}
\caption{Quiver associated to the D1-D3-NS5 brane subsystem for $r$ constant. Circles denote (4,4) vector multiplets and black lines (4,4) bifundamental hypermultiplets. The gauge groups with ranks $p\alpha_k$, with $k=1,\dots, P-1$ to the left of the gauge group with rank $p\alpha_P$ have U(1) flavour symmetries. The gauge group with rank $p\alpha_P$ has U(2) flavour symmetry. The rest of gauge groups do not have attached any flavours. We have used units of $q$.}
\label{quiverD1D3}
\end{figure}

\begin{figure}
\centering
\includegraphics[scale=0.72]{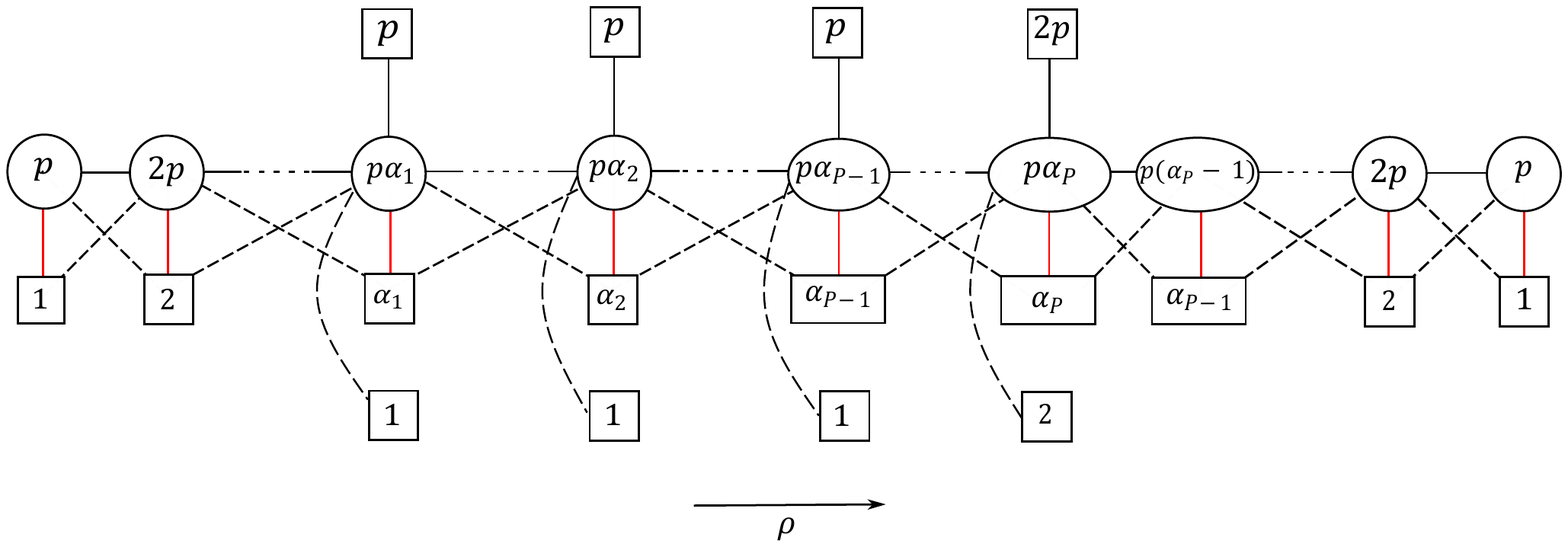}
\vspace{-16.5cm}
\caption{Quiver associated to the D1-D3-D5-D7-NS5 brane system for $r$ constant. Circles denote (4,4) vector multiplets, black lines (4,4) bifundamental hypermultiplets, red lines (0,4) bifundamental twisted hypermultiplets and dashed lines bifundamental (0,2) Fermi multiplets. The gauge groups associated to the D5-branes have become flavour groups since these branes are much heavier than the D1-branes. One can check that all gauge nodes are balanced (ie $2N_c=N_f$, with $N_c$ the number of colours and $N_f$ the number of flavours) but the ones with ranks $p\alpha_k$, with $k=1,2,\dots, P-1$ to the left of the gauge group with rank $p\alpha_P$. We have used units of $q$.}
\label{finalquiver}
\end{figure}

The results in Appendix \ref{fieldtheory} show that the D5-D7-NS5 and D1-D3-NS5 subsystems of the brane set-up are described by the (4,4) quivers depicted in Figures \ref{quiverD5D7} and \ref{quiverD1D3}. 
Our next step is to couple these two quivers to each other. This will reduce the supersymmetries to (0,4). The new massless modes that arise are the ones associated to the open strings stretched between the D1 and the D5 and D7 branes, given by:
\begin{itemize}
\item D1-D5 strings: Strings with one end on D1-branes and the other end on orthogonal D5-branes in the same interval between NS5-branes contribute with fundamental (0,4) twisted hypermultiplets, associated to the motion of the string along the $(x^6,x^7,x^8,x^9)$ directions, which are charged under the R-symmetry. Strings with one end on D1-branes and the other end on D5-branes in adjacent intervals between NS5-branes contribute with fundamental (0,2) Fermi multiplets, since all the scalars are fixed.
 \item D1-D7 strings: Strings with one end on D1-branes and the other end on D7-branes in the same interval contribute with fundamental (0,2) Fermi multiplets. This can be seeing more concretely by noting that the set-up is T-dual to the D0-D8 system. 
\end{itemize}
These new massless modes thus render the quiver (0,4) (or $\mathcal{N}=4$ in 1d) supersymmetric. The resulting quiver is depicted in Figure \ref{finalquiver}. 
Our proposal is that this quiver describes a supersymmetric quantum mechanics that flows in the IR to the SCQM dual to our solutions. Similar quivers to the ones constructed in this subsection have been proposed in the literature in order to describe the bubbling sector of singular 't Hooft monopoles in 4d $\mathcal{N}=2$ theories living in D3-D7 branes, in particular in the presence of D5-branes (see \cite{Assel:2019iae}). We will come back to this discussion in subsection \ref{thooft}. Previous to that we will check our proposal with the computation of the central charge.

 \subsubsection{Computation of the central charge}
 
 As a check of our proposal we proceed now to the computation of the field theory and holographic central charges. We start with the field theory calculation. 
 
It is well-known that an $\mathcal{N}=4$ quantum mechanics can be described in terms of 2d $(0,4)$ multiplets, and that both theories share the same superconformal algebra. Therefore, we can use that the superconformal algebra relates the (right-moving) central charge to the R-symmetry anomaly (the level of the superconformal R-symmetry) to compute the central charge also in one dimension. The relation is that 
 \begin{equation}
 c_R=6(n_{hyp}-n_{vec}),
 \end{equation}
 where $n_{hyp}$ stands for the number of (0,4) untwisted hypermultiplets and $n_{vec}$ for the number of (0,4) vector multiplets. Indeed, this expression has been successfully\footnote{In the sense that it agrees with the corresponding holographic results.} used in previous computations of the central charge of $\mathcal{N}=4$ SCQMs (see \cite{Lozano:2020txg,Lozano:2020sae,Lozano:2021rmk,Ramirez:2021tkd,Lozano:2021fkk,Lozano:2022vsv}). 
 For the quiver depicted in Figure \ref{finalquiver} this gives, 
 \begin{equation}
 n_{hyp}^{(n)}=q^2p^2\, \Bigl(2\sum_{k=1}^{\alpha_P-1} k(k+1)+\sum_{k=1}^{P-1}\alpha_k+2\alpha_P\Bigr) 
 \end{equation}
 and
 \begin{equation}
 n_{vec}^{(n)}=q^2p^2\, \Bigl(2\sum_{k=1}^{\alpha_P-1} k^2 +\alpha_P^2\Bigr),
 \end{equation}
 with $p=(n+\frac12)$, and finally
 \begin{equation}
 c_R^{(n)}= 6 q^2 p^2\, \sum_{k=1}^{P} \alpha_k.
 \end{equation}
This gives the contribution to the central charge of the $r\in [2n\pi,2(n+1)\pi]$ interval with $n=1,2,\dots, P'$. Adding these contributions plus the ones with $n=P'+1,P'+2,\dots, 2P'$, for which $p\rightarrow 2(P'+1)-p$, we find
\begin{equation} \label{fieldtheorycR}
c_R=q^2\Bigl(4P'^3+12P'^2+11P'\Bigr)\sum_{k=1}^{P} \alpha_k\sim 4q^2 P'^3 \sum_{k=1}^{P} \alpha_k
\end{equation}
to leading order in $P'$. 

Let us proceed now with the computation of the holographic central charge. We use that \cite{Lozano:2020txg}
\begin{equation}
c_{hol}=\frac{3}{4\pi G_N}\int d^8\xi \,e^{-2\Phi}\sqrt{\text{det} g_8},
\end{equation}
which gives for our solutions
\begin{equation}
c_{hol}=q^2 \Bigl(4P'^3+12P'^2\Bigr)  \sum_{k=1}^P \alpha_k\sim 4q^2 P'^3 \sum_{k=1}^P \alpha_k,
\end{equation}
to leading order in $P'$. In fact, the quantity that needs to be compared to the holographic central charge is 
\begin{equation}
c_{hol} \longleftrightarrow \frac{c_L+c_R}{2}
\end{equation}
with 
\begin{equation}
c_L=c_R+\text{Tr}\gamma^3,
\end{equation}
where the trace is over the Weyl fermions in the theory and $\gamma^3$ is the chirality matrix in 2d. The (4,4) multiplets in the quiver quantum mechanics depicted in Figure \ref{finalquiver} contain the same number of left-handed and right-handed multiplets.  Therefore, the only ones that contribute to the previous expression are the ones in (0,4) multiplets, right-handed, and the ones in (0,2) Fermi multiplets, left-handed. We obtain
\begin{equation}
 \text{Tr}\gamma^3=-2 q^2 \sum_{n=1}^{P'}(n+\frac12)\sum_{k=1}^{P-1}\alpha_k=-q^2 P'(P'+2)\sum_{k=1}^{P-1}\alpha_k.
 \end{equation} 
 Putting this result together with the field theory computation \eqref{fieldtheorycR} we get
 \begin{equation}
 \frac{c_L+c_R}{2}=q^2 \Bigl( 4P'^3+\frac{23}{2}P'^2+10P'\Bigr)\sum_{k=1}^P \alpha_k+\frac12 q^2 P'(P'+2)\alpha_P\sim 4q^2 P'^3 \sum_{k=1}^P \alpha_k,
 \end{equation}
 to leading order in $P'$. Therefore, we find perfect agreement between the field theory and holographic results, to leading order in $P'$. 

\subsection{'t Hooft defect interpretation}\label{thooft}

Quiver quantum mechanics like the ones discussed in the previous subsection have been used in the literature in localization computations of monopole bubbling contributions \cite{Kapustin:2006pk} to supersymmetric 't Hooft loops, in 4d $\mathcal{N}=2$  supersymmetric gauge theories \cite{Brennan:2018yuj,Brennan:2018moe,Brennan:2018rcn,Assel:2019iae}. In these calculations the contribution of the monopole bubbling sector is computed as the index of a supersymmetric quantum mechanics, defined from  the string theory realisation of the 't Hooft loop bubbling sector. 

\begin{figure}
\centering
\includegraphics[scale=0.55]{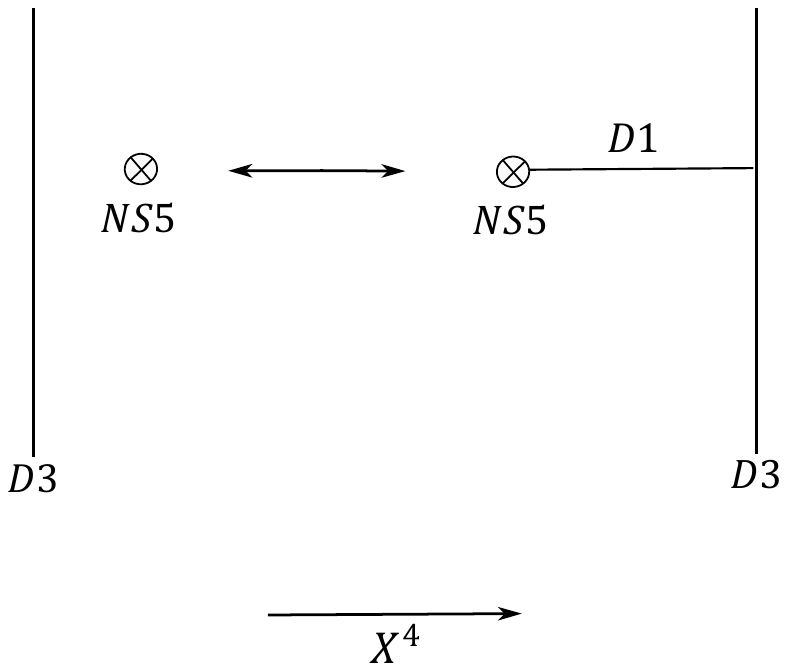}
\vspace{-11.5cm}
\caption{Brane realisation of singular 't Hooft monopoles in 4d D3-brane systems.}
\label{thooftloop}
\end{figure}

In 4d D3-D7 systems smooth monopoles are realised by D1-branes stretched between the D3-branes, according to the Nahm construction \cite{Diaconescu:1996rk}. In turn, singular 't Hooft monopoles are realised by introducing spatially transverse NS5-branes \cite{Cherkis:1997aa,Brennan:2018moe}, which source $\pm \frac12$ units of magnetic charge in the worldvolume of the D3-branes (see Figure \ref{thooftloop}). 
Since these D1-branes have Dirichlet and Neumann boundary conditions at opposite ends they do not carry any degrees of freedom, as expected from 't Hooft operators.  Monopole bubbling occurs when the smooth monopoles are absorbed by the 't Hooft defect, decreasing its effective magnetic charge  \cite{Cherkis:1997aa,Brennan:2018moe}. In the brane setting this occurs when D1-branes stretched between adjacent D3-branes coincide with NS5-branes \cite{Brennan:2018moe,Brennan:2018yuj}. The SQM from which the bubbling sector is computed arises then as the low energy theory on the D1-branes of the brane set-up, depicted in Table \ref{D3D7NS5D1}. 
 \begin{table}[http!]
	\begin{center}
		\begin{tabular}{| l | c | c | c| c | c | c | c | c| c | c |}
			\hline		    
			& $t$ & $x^1$  & $x^2$ & $x^3$ & $x^4$ & $x^5$ & $x^6$ & $x^7$ & $x^8$ & $x^9$ \\ \hline
			D3 & x & x & x & x &  & & & & & \\ \hline
			D7 & x & x & x & x &  & &x &x &x &x \\ \hline
			NS5 & x &  &  &  &  &x &x  & x  & x & x  \\ \hline
			D1 & x &  &  & & x &  & & & & \\ \hline
		\end{tabular} 
	\end{center}
	\caption{Brane set-up for the bubbling sector of 4d $\mathcal{N}=2$ D3-D7 theories.} 
		\label{D3D7NS5D1}	
\end{table}
This has been supported by localization computations that show that the monopole bubbling contribution to the partition function of the 4d theory is obtained as the Witten index of the SQM living in the D1-branes \cite{Brennan:2018yuj,Brennan:2018rcn,Assel:2019iae}.

Interestingly, it was shown in \cite{Assel:2019iae} that in order to correctly account for the bending of the NS5-branes caused by the branch cut of the D7-branes, extra D5-branes need to be introduced in the previous brane set-up, along with F1-strings. The D5-branes introduce new fields in the SQM, coming from the D1-D5 open strings, that also contribute to the Witten index. The introduction of the D5 and F1 branes renders the whole brane intersection realising the SQM S-duality symmetric, such that it can describe at the same time the bubbling of 't Hooft loops, associated to the NS5-D1 subsector, of Wilson loops, coming from the D5-F1 branes, or of dyonic loops. It was in fact argued in \cite{Assel:2019iae} that the bubbling sector computed from the SQM is associated to dyonic loops, given the presence of both D1 and F1-strings in the brane configuration. The brane set-up that arises when the D5-branes and the F1-strings are added is the one depicted in
Table \ref{D3D7NS5D1mod}, which is no other than the brane intersection proposed in Table \ref{D1D7NS5D5D3F1} as underlying our $\text{AdS}_2$ solutions.
  \begin{table}[http!]
	\begin{center}
		\begin{tabular}{| l | c | c | c| c | c | c | c | c| c | c |}
			\hline		    
			& $x^0$ & $x^1$  & $x^2$ & $x^3$ & $x^4$ & $x^5$ & $x^6$ & $x^7$ & $x^8$ & $x^9$ \\ \hline
			D3 & x & x & x & x &  & & & & & \\ \hline
			D7 & x & x & x & x &  & &x &x &x &x \\ \hline
			NS5 & x &   &  &  &   &x &x  & x  & x & x  \\ \hline
			D1 & x &  &  & & x&  & & & & \\ \hline
			D5 & x & & & &x &  & x & x & x & x \\ \hline
			F1 & x & & & & & x & & & &  \\ \hline
		\end{tabular} 
	\end{center}
	\caption{Brane set-up for the bubbling sector of 4d $\mathcal{N}=2$ D3-D7 theories including D5 and F1 branes \cite{Assel:2019iae}.} 
		\label{D3D7NS5D1mod}	
\end{table}
 It is thus natural to expect that our solutions may provide a geometrical description of the bubbling of dyonic loops in 4d $\mathcal{N}=2$ D3-D7 systems. 

Further to this, it was shown in \cite{Brennan:2018yuj} that the bubbling sector of 't Hooft loops in the 4d $\mathcal{N}=2$ SU(N) theory living in D3-branes in an Omega background is described by quiver quantum mechanics consisting on N-1 balanced quivers\footnote{That satisfy that $N_f=2N_c$, with $N_f$ the number of flavours that couple to the gauge node and $N_c$ its rank.} of length the number of NS5-branes in each interval between two D3-branes, connected by N-2 unbalanced gauge nodes. The N D3-branes of the 4d theory contribute to these quivers with flavour groups, that couple to the unbalanced gauge nodes. The detailed structure of the unbalanced quivers contains the information about the particular bubbling sector described by the quiver QM, as carefully explained in  \cite{Brennan:2018yuj}. These constructions were then extended in \cite{Brennan:2018rcn} to include $N_f$ D7-branes in the 4d SU(N) gauge theory.
Interestingly, in reference \cite{Assel:2019iae} these quivers were improved to account for the bending of the NS5-branes caused by the D7-branes. In this completed brane set-up the D3 and D7 branes are interpreted as sitting inside a 5-brane web. The resulting QM living in the completed brane set-up is a (0,4) supersymmetric quiver theory built out of (4,4) vector multiplets coming from each gauge node, (4,4) bifundamentals coming from D1-D1 branes across NS5-branes, twisted (0,4) bifundamentals coming from D1-D5 strings ((0,2) bifundamentals if across an NS5-brane) and (0,2) bifundamental Fermi multiplets coming from D1-D7 strings. This is exactly the field content of the quiver quantum mechanics depicted in Figure \ref{finalquiver}, which can thus be interpreted as describing dyonic loops in 4d $\mathcal{N}=2$ theories living in D3-D7 branes. 

Let us analyse in a bit more detail this interpretation. In the bubbling description the number of NS5-branes between stacks of D3-branes has to be an even integer number, for them to induce an integer magnetic charge in the worldvolume of the D3-branes (see for instance \cite{Assel:2019iae}). This fixes $\beta_k\in 2\mathbb{Z}$, $p\alpha_k\in \mathbb{Z}$, and therefore the number of D1-branes in each interval to an integer number.  Second, we can identify the precise bubbling sector that our quiver quantum mechanics is describing. For this we can follow the analysis in \cite{Brennan:2018yuj} for the D1-D3-NS5 brane subsystem, where the information about the bubbling sector is encoded. 
Comparison with the quivers proposed in \cite{Brennan:2018yuj} shows that our quiver in Figure \ref{finalquiver} would describe the bubbling sector in which all the magnetic charge of the defect has been screened\footnote{The sector with ${\bf v}=\text{diag}(0,\dots, 0)$ in the notation of \cite{Brennan:2018yuj}.}, of the 4d SU($qp(P+1)$) gauge theory living in the D3-branes, with $N_f=q (P+1)$ flavour groups coming from the D7-branes, in a given $r\in [2n\pi,2(n+1)\pi]$ interval. One can check that our quiver consists indeed of $P$ balanced sub-quivers of length $\beta_k$, with $k=0,1,\dots, \beta_{P-1}$, separated by $P-1$ unbalanced ones\footnote{Note that our quivers are scaled by a factor $qp$ compared to the ones in \cite{Brennan:2018yuj}.}, to which the flavour groups introduced by the D3-D7 strings couple. Note that in our quantum mechanics  the condition for a gauge node to be balanced is inherited from the 2d theory, where the cancellation of the gauge anomaly imposes that the number of (0,2) Fermi multiplets connected to a given gauge node must be twice the number of (0,4) hypermultiplets that couple to this node.

A related question to address is whether the 4d theory living in the D3-D7 branes is conformal. Our construction gives $N_c= qp (P+1)= (n+\frac12) N_f$, and therefore differs from the condition $N_c=\frac12 N_f$ required by conformal invariance. However, one can note that it is possible to cancel the term proportional to $n$ by adding a worlvolume flux with instanton number $\int_{\mathbb{R}^4}F\wedge F=8\pi^2 n$, that would render the 4d theory conformally invariant. 

\begin{figure}
\centering
\includegraphics[scale=0.5]{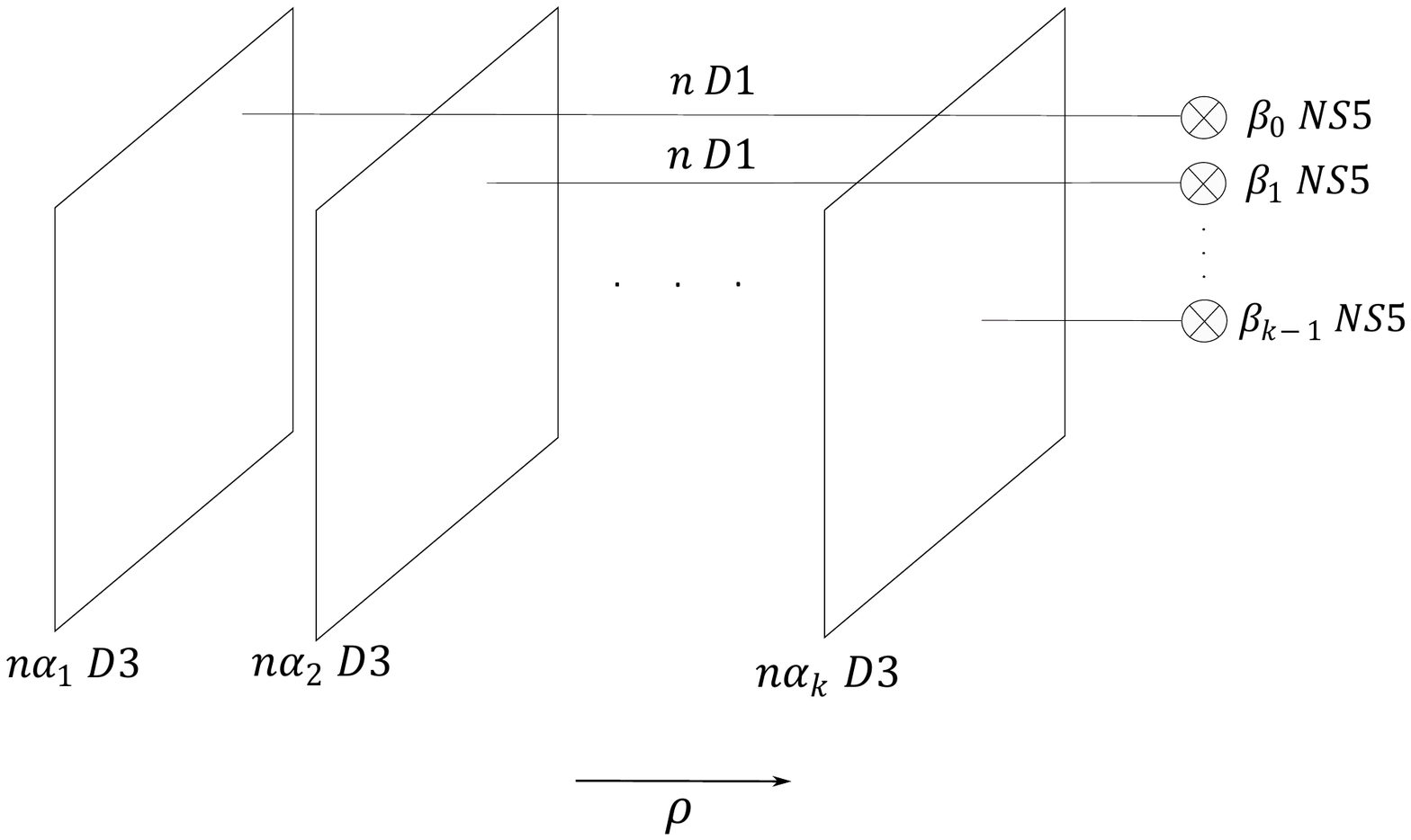}
\vspace{-9cm}
\caption{D1-D3-NS5 sector of the brane set-up depicted in Figure \ref{brane-set-up-rho} after a sequence of Hanany-Witten moves, where in this case we use electric instead of magnetic charges for the D3 and D1 branes.
}
\label{thooftmonopole}
\end{figure}

The interpretation of the D1-D3-NS5 subsystem of the brane set-up as associated to 't Hooft loops becomes clearer when one notices that this part of the brane set-up is just the S-dual of the F1-D3-D5 subsector used to describe Wilson loops (or baryon vertices). Indeed, dualising the Born-Infeld vector field in the worldvolume of the D3-branes to its electric-magnetic dual $\tilde A$, one finds a coupling
\begin{equation}
S_{D3}=T_3\int B \wedge d{\tilde A}=-T_3 \int H\wedge {\tilde A},
\end{equation}
that shows that a D1-string with electric charge $nq$ induces $nq \beta_k$ D3-brane electric charge. Adding the contributions of all $\rho$-intervals one finds a total $Q_{D3}^e=nq \alpha_k$ electric charge, as found in \eqref{QD3e}. This is illustrated in Figure \ref{thooftmonopole}, which can be obtained from the D1-D3-NS5 sector of the brane set-up depicted in Figure \ref{brane-set-up-rho} after a sequence of Hanany-Witten moves, where in this case we use electric instead of magnetic charges to count the numbers of D3 and D1 branes.
The D3-branes thus find a dual interpretation: as baryon vertices together with the F1 and the D5 branes, in the $r$-space, and as 't Hooft lines together with the D1 and the NS5-branes in the $\rho$-space. The 't Hooft line is again in the completely antisymmetric representation of  U($qn\alpha_1$)$\times$  U($qn\alpha_2$) $\times \dots $ U($qn\alpha_P)$, and could thus be interpreted as a sort of magnetic baryon vertex. The reader can check the similarity between Figure \ref{thooftmonopole} and the D3-F1-NS5 sector of Figure \ref{HW-moves}, up to the different global completions used for the $\rho$ and $r$ directions. 

Our analysis in this subsection suggests an interpretation of the $\text{AdS}_2$ solutions constructed in section \ref{eq:case2SL2Tdual} as geometric duals of SCQMs describing the bubbling sector of vanishing effective magnetic charge in 4d $\mathcal{N}=2$ SCFTs living in D3-D7 systems. It would be interesting to investigate whether the broader solutions constructed  in section \ref{sec:genclass} could geometrically describe more general bubbling sectors. Note that based on this interpretation one would expect  
 that the solutions \eqref{eq:4.1}-\eqref{eq:4.1b} asymptoted locally to the $\text{AdS}_5$ spacetime dual to the 4d SCFT living in the D3-D7 branes. This is however not the case for our class of solutions. Indeed, in the two different interpretations that we have discussed the D3-branes either play the role of background branes, in their interpretation in this section, or defect branes, in the baryon vertex interpretation discussed in  section \ref{baryon-vertex-interpretation}.
Therefore, compared to the simpler configurations studied in \cite{Lozano:2020sae,Lozano:2021rmk,Lozano:2021fkk,Lozano:2022vsv}, in this case there is no clear distinction between defect and background branes, and one should not expect that a higher dimensional AdS space, that is typically identified as the near horizon geometry of the background branes, arises asymptotically locally from the solutions, in agreement with our findings. It would be interesting to construct the explicit 
 brane intersection that underlies the solutions to shed further light on their possible defect interpretation.

Related to our previous observation, it is worth mentioning that  the brane intersection given in Table \ref{D1D7NS5D5D3F1} can more generally be regarded as a coupled 5d-4d-1d system, from which one can compute the bubbling sectors of 't Hooft loops in the 4d theory living in the D3-D7 branes but also the instanton sectors of the 5d theory living in the 5-brane web. The latter approach was the one taken in \cite{Assel:2018rcw}, where the same brane set-up depicted in Table \ref{D1D7NS5D5D3F1} was proposed for the study of loop operators in the 5d theory living on D5-NS5 branes. In this setting the D3-branes introduce Wilson lines, together with the F1-strings, and S-dual Wilson lines, together with the D1-branes, that in this case connect the D3-branes with the NS5-branes of the 5-brane web. Note that the condition for the 5d theory to flow to a conformal point in the UV, given by $2N_c\ge N_f+c_{cl}$ is satisfied by the D5-D7-NS5 subsector of our quiver mechanics. Connecting to our discussion in the previous paragraph, it is worth stressing that in this interpretation the D3-branes would play the role of defect branes in both the instanton and baryon vertex interpretations, with the D5-D7-NS5 branes playing the role of background branes also in both interpretations. Thus, one could expect that the AdS$_6$ geometry arising in the near horizon of the 5-brane web may emerge asymptotically locally from the solutions. As we discussed in section \ref{baryon-vertex-interpretation} this is however not the case. As mentioned above the explicit construction of the brane intersection underlying the solutions would be very enlightening in clarifying this issue.

\section{Conclusions}

In this paper we have presented general results on the generation of $\text{AdS}_2$ solutions to Type II supergravities from $\text{AdS}_3$ backgrounds using U(1) and SL(2) T-dualities, paying special attention to the conditions for preservation of supersymmetry. We have discussed in detail the solutions that arise from the $\text{AdS}_3\times \text{S}^3$ solutions with $\mathcal{N}=(0,4)$ supersymmetries and SU(3) structure studied in \cite{Lozano:2022ouq}, in particular to the ones constructed via SL(2) T-duality with $h$ constant. We have shown that these solutions belong to the more general class obtained from the $\text{AdS}_3\times S^2$ solutions to Type IIB supergravity constructed in \cite{Macpherson:2022sbs} through a double analytical continuation. 

We have initiated the study of the field theory duals to the new $\text{AdS}_2$ solutions focusing on the previously alluded to SL(2) T-dual, consisting on $\text{AdS}_2\times \text{S}^3\times \mathbb{T}^3$ foliations over a 2d Riemann surface, parametrised by two non-compact directions $(\rho,r)$. We have proposed that these solutions are dual to SCQMs arising in the IR from supersymmetric quantum mechanics living in D1-branes extended in the $\rho$ direction between NS5-branes, with flavour groups arising from D3, D5 and D7 branes. We have seen that the F1-strings also present in the brane set-up play the role,  
together with the D3-branes, of baryon vertices for the D5-branes, and, together with the D1-branes, of baryon vertices for the D7-branes. This structure is revealed by looking at the position of the branes along the $r$ direction. The D3-D5-F1 and D1-D7-F1 brane subsystems are indeed displayed exactly as in the D3-D5-F1 brane configurations studied in  \cite{Yamaguchi:2006tq,Gomis:2006sb}, describing Wilson lines in antisymmetric representations in 4d $\mathcal{N}=4$ SYM. As encountered in previous examples of $\text{AdS}_2$ solutions associated to baryon vertex configurations \cite{Lozano:2020sae,Lozano:2021rmk,Ramirez:2021tkd,Lozano:2021fkk,Lozano:2022vsv}, one can associate the solutions to D5-D7-NS5 brane intersections, where a 5d $\mathcal{N}=1$ theory lives, in which one dimensional defects are introduced. In the IR the gauge symmetry on the D5-D7-NS5 system becomes global, turning them from colour to flavour branes, with the defect branes becoming the new colour branes of the backreacted geometry. In our construction in this paper, however, the presence of the second non-compact direction in the geometry (not present in the examples in the previous references), playing the role of  field theory direction along which the D1-branes extend, renders the D1-branes the only remaining colour branes of the configuration.  

The intricate brane set-up associated to the solutions unveils, moreover, a similar structure along the $\rho$-direction, to the one arising along the $r$-direction, with the D1-branes playing the same role as the F1-strings now in conjunction with the D3-branes and the NS5-branes,  to describe, in this case, 't Hooft monopoles in the 4d theory living on D3-D7 branes.  Quiver quantum mechanics like the ones we have explicitly constructed have been proposed in the literature in the description of monopole bubbling in 4d $\mathcal{N}=2$ supersymmetric theories living in these 4d intersections \cite{Brennan:2018yuj,Brennan:2018moe,Brennan:2018rcn,Assel:2019iae}. These constructions involve the D1-D3-NS5-D7 subsystem of the brane set-up associated to the solutions. Interestingly, in order to account for the bending of the NS5-branes due to the branch cuts introduced by the D7-branes it was shown in \cite{Assel:2019iae} that D5-branes (and F1-strings) need also be introduced in the description, such that the final brane set-up must be precisely the one underlying our solutions. Our $\text{AdS}_2$ solutions thus find a natural interpretation as geometrical descriptions of monopole bubbling in 4d $\mathcal{N}=2$ D3-D7 systems. As discussed in the paper, the bubbling sector described by our quantum mechanics would be the one of vanishing effective magnetic charge. It is likely that the more general solutions presented in section 3 would provide the geometrical description of more general bubbling sectors. It would be interesting to investigate this further.

\section*{Acknowledgments}
We are supported by grants
from the Spanish government MCIU-22-PID2021-123021NB-I00 and Principality of Asturias SV-PA-21-AYUD/2021/52177. The work of NM is also supported by the Ram\'on y Cajal fellowship RYC2021-033794-I.

\appendix

\section{Conventions}\label{sec:appendixonconvensions}
We follow democratic like conventions of \cite{Tomasiello:2022dwe}  for Type II.  

The bosonic sector of Type II supergravity consists of the NS-NS and RR sectors. The former containing the metric, dilaton $\Phi$ and NS 3-form $H$. The later can be expressed in term of a polyform $F$ such that
\beq
F= \left\{\begin{array}{l} F_0+F_2+F_4+F_6+F_8+F_{10}~~~~\text{IIA}\\[2mm] F_1+F_3+F_5+F_7+F_9~~~~\text{IIB}\end{array}\right.
\eeq
The polyform is subject to a self duality constraint which halfs its degrees of freedom, namely
\beq
F=\star \lambda(F),
\eeq
where $\lambda (C_k)= (-1)^{[\frac{k}{2}]}C_k$. The Hodge dual is defined in terms of the $d=10$ vielbein $e^{\underline{M}}$ as
\beq
\star e^{\underline{M}_1...\underline{M}_k}=\frac{1}{(d-k)!}\epsilon_{\underline{M}_{k+1}...\underline{M}_{d-k}}^{~~~~~~~~~~~~~\underline{M}_1...\underline{M}_k}e^{\underline{M}_{k+1}...\underline{M}_{d-k}},
\eeq
where the $d=10$ indices $M$ are curved  and $\underline{M}$ are flat. 

A solution to type II supergravity is one that solves the following equations of motion
\begin{align}
&d_H F=0,~~~~ dH=0,~~~~ d(e^{-2\Phi}\star_{10}H)-\frac{1}{2}(F,F)_8=0,
\nn\\[2mm]
& 2R^{(10)}- H^2-8 e^{\Phi}(\nabla^{(10)})^2 e^{-\Phi}=0,~~~  R^{(10)}_{AB}+2 \nabla^{(10)}_{A}\nabla^{(10)}_{B}\Phi-\frac{1}{2} H^2_{AB}-\frac{e^{\Phi}}{4} (F)^2_{AB}=0,\label{eq:EOM}
\end{align}
where $(F,F)_8$ is the 8-form part of $F\wedge \lambda(F)$ and we refer the reader to \cite{Tomasiello:2022dwe}  for further details. In the presence of sources these equations get modified.

A solution of Type II supergravity preserves supersymmetry if the following spinorial conditions can be solved for non trivial $d=10$ Majorana-Weyl spinors $\epsilon_{1,2}$
\begin{subequations}
\begin{align}
& \left(\nabla^{(10)}_M-\frac{1}{4}H_M\right) \epsilon_1 + \frac{e^\Phi}{16}F \Gamma_M \epsilon^2=0, \label{eq:10dsusyeqs1}\\[2mm]
& \left(\nabla^{(10)}_M+\frac{1}{4}H_M\right) \epsilon_2 \pm \frac{e^\Phi}{16} \lambda(F) \Gamma_M \epsilon^1=0, \label{eq:10dsusyeqs2}\\[2mm]
& \left(\nabla^{(10)}-\frac{1}{4}H-d\Phi\right)\epsilon_1=0, \label{eq:10dsusyeqs3} \\[2mm]
& \left(\nabla^{(10)}+\frac{1}{4}H-d\Phi\right)\epsilon_2=0,\label{eq:10dsusyeqs4}
\end{align}
\end{subequations}
where the Clifford map is assumed and we define the spin covariant derivative as
\beq
\nabla_M= \partial_{M}+ \frac{1}{4}\omega_M{}^{\underline{P}\underline{Q}}\Gamma_{\underline{P}\underline{Q}},~~~~ d e^{\underline{M}}+\omega^{\underline{M}}{}_{\underline{N}}\wedge e^{\underline{N}}=0.
\eeq
The upper/lower signs are taken in IIA/IIB and  for the chirality matrix defined as $\hat \Gamma =  \Gamma^{\underline{0}...\underline{9}}$ we have that
\beq
\hat \Gamma \epsilon_1= \epsilon_1,~~~~\hat \Gamma \epsilon_2=\mp \epsilon_2.
\eeq
We also have the useful identities
\beq
\hat \Gamma C_{k}= \star \lambda(C_k),~~~~~\hat\Gamma C_k =- \lambda(\star C_k).\label{eq:usefulid}
\eeq
Throughout the next sections we shall make use of a 3+7 split of the $d=10$ gamma matrices such that  
\begin{align}
\Gamma_{\underline{\mu}} & = \gamma_{\underline{\mu}} \otimes \sigma_3 \otimes \mathbbm{1}_8, \qquad & \mu={0,1,2}, \nn \\[2mm]
\Gamma_a & = \mathbbm{1}_2 \otimes \sigma_1 \otimes \gamma_a, \qquad  & a={1,..,7},~~~~i\gamma_{1...7}=\mathbbm{1}_8\label{eq:10dgammas}
\end{align}
where $\gamma_{\underline{\mu}}= (i\sigma_2,\sigma_1,\sigma_3)_{\underline{\mu}}$ and are thus real. 
The $d=10$ intertwiner defining Majorana conjugation as $\epsilon^c= B^{(10)}\epsilon^*$ and chirality matrix are then
\beq
B^{(10)}=\mathbbm{1}_2\otimes \sigma_3\otimes B,~~~~ B^{-1}\gamma_a B=-\gamma_a^{*},~~~~\hat \Gamma= -\mathbbm{1}_2 \otimes \sigma_2\otimes \mathbbm{1}_8
\eeq

\section{AdS$_3$, its isometries and Killing spinors}\label{sec:detailsonAdS3}
In this appendix we explore some details of the isometries of AdS$_3$, the Killing vectors it supports and what they are charged under. This will be used in the following appendices.\\
~\\
One can express the Hopf fibration of unit radius AdS$_3$ as
\beq
ds^2(\text{AdS}_3)= \frac{1}{4}ds^2(\text{AdS}_2)+\frac{1}{4} (dr+V)^2,~~~~ dV= s \text{vol}(\text{AdS}_2),~~~~s^2=1,
\eeq
where $\partial_r$ is an isometry. 
A useful parameterisation for what follows is to take
\beq
ds^2(\text{AdS}_2)= -\cosh^2 x dt^2+ dx^2,~~~V= -s \sinh x dt.
\eeq
AdS$_3$ has a global SO(2,2) symmetry with 2 distinct SL(2) subgroups we shall label as SL(2)$_{L/R}$ with $L/R$ standing for left and right.  We define the generators of $\mathfrak{sl}(2)$ to be
\beq
\tau_i=\frac{1}{2}(i\sigma_2,\sigma_1,\sigma_3)_i,
\eeq
for $\sigma_{1,2,3}$ the Pauli matrices. These obey the Lie bracket relation
\beq
[\tau_i,\tau_j]=f_{ij}^{~~k}\tau_k,
\eeq
where the structure constants can be easily computed and indices are raised and lowered with the standard mostly positive Minkowki$_3$ metric $\eta=\text{Diag}(-1,1,1)$.  We define an element of SL(2) to be
\beq
G= e^{- s t\tau_1}e^{-  s x\tau_2} e^{-  s r\tau_3},
\eeq
with left/right invariant 1-forms on AdS$_3$ defined in terms of this as
\beq
L^i= -2 s\text{Tr}\bigg[\tau^i G^{-1} d G\bigg],~~~R^i= -2 s\text{Tr}\bigg[\tau^i d G G^{-1} \bigg],
\eeq
which both span AdS$_3$ as
\beq
ds^2(\text{AdS}_3)=  \frac{1}{4}L_i L^i= \frac{1}{4} R_i R^i,
\eeq
and  obey the following differential relation
\beq
dL^i= \frac{s}{2}f^i_{~jk}L^j\wedge L^k,~~~~dR^i= -\frac{s}{2}f^i_{~jk}R^j\wedge R^k.
\eeq
The vectors dual to the left/right invariant forms define the SL(2)$_{R/L}$ Killing vectors as
\beq
K_{R/L}^i = \frac{1}{4} (L/R)^i_{\mu}  (g_{\text{AdS}_3})^{\mu\nu}\partial_{\mu}.
\eeq
with the factor of $\frac{1}{4}$ taken such that $(K_{L/R}^i)^{\mu} ((R/L)^j)_{\mu}=\eta^{ij}$, we also have the relations
\beq
{\cal L}_{K^i_{L/R}} (L/R)^j=0,~~~~ {\cal L}_{K^i_{L/R}} (R/L)^j =( +/-)f^{ij}_{~~k} (R/L)^k.
\eeq 
With respect to our specific parameterisation the Killing vectors are
\begin{align}
K_{R}^1\pm K_{R}^2&= e^{\pm s r}(s \tanh x \partial_r+\sech x \partial_t-\partial_x),~~~~ K_{R}^3=\partial_r,\nn\\[2mm]
K_{L}^3+i K_{L}^2&= e^{i s t}(\sech x\partial_r- s \tanh x \partial t+i \partial_x),~~~~K_{L}^1=-\partial_t,
\end{align}
so in particular it SL(2)$_R$ that includes the U(1) isometry corresponding to $\partial_r$.

AdS$_3$ supports two physically distinct types of spinors that transform as doublets with respect to one of SL(2)$_{L/R}$ whilst also being singlets with respect to SL(2)$_{R/L}$. To see this let us define the following frame maintaining the Hopf fibration structure,
\beq
e^{\underline{\mu}}_3= \frac{1}{2}( \cosh x dt,dx,(dr+V))^{\underline{\mu}},\label{eq:AdS3frame}
\eeq
where an underscore indicates that an index is flat.
We shall define the independent spinors on (unit radius)  AdS$_3$ to solve the Killing spinor equations
\beq
\nabla_{\mu} \zeta_{R/L}=(+/-) \frac{s}{2}\gamma_{\mu} \zeta_{R/L}. \label{eq:LRKSE}
\eeq
One can show with respect to the above frame that these constraints are solved by
\beq\label{eq:AdS3spinors}
\zeta_R=   e^{\frac{s}{2}x\sigma_1}e^{\frac{s}{2}t i \sigma_2}\zeta_R^0,~~~~ \zeta_L=   e^{-\frac{s}{2}r \sigma_3} \zeta_L^0,
\eeq
where $\zeta^0_{R/L}$ are arbitrary constant spinors, they need only be real for  $\zeta_{R/L}$ to be Majorana in our conventions.  In terms of these we can define our spinor doublets:  Defining unit norm constant spinors $\zeta^0_{\pm}$ such that $\sigma_3\zeta^0_{\pm}=\pm \zeta^0_{\pm}$ we have that
\beq
\zeta^A_R=\left(\begin{array}{c}e^{\frac{s}{2}x\sigma_1}e^{\frac{s}{2}t i \sigma_2}\zeta^0_{+}\\e^{\frac{s}{2}x\sigma_1}e^{\frac{s}{2}t i \sigma_2}\zeta^0_{-}\end{array}\right)^A,~~~~~\zeta^A_L=\left(\begin{array}{c}e^{-\frac{s}{2}r \sigma_3}\zeta^0_{+}\\e^{-\frac{s}{2}r \sigma_3}\zeta^0_{-}\end{array}\right)^A
\eeq
are such that
\beq
{\cal L}_{K^i_{R/L}} \zeta^A_{R/L}=0,~~~~~{\cal L}_{K^i_{R/L}} \zeta^A_{L/R}=(-/+) s (\tau_i)^A_{~B}\zeta^B_{L/R}.
\eeq
We thus find that it is $\zeta_R$ that is a singlet with respect to the isometry $\partial_r$, and likewise the rest of SL(2)$_R$, spanned by $K^i_{R}$. An important thing to also appreciate is that the Killing spinor $\zeta_R$ additionally solves the Killing spinor equation on round AdS$_2$, namely
\beq
\nabla_{\mu_2}^{\text{AdS}_2}\zeta_R= \frac{s}{2} \gamma_{\mu_2}\zeta_R,
\eeq
when the frame is $e^{\underline{\mu}_2}_2= (\cosh x dt,dx)^{\underline{\mu}_2}$ and $\underline{\mu}=(\underline{\mu}_2,\underline{r})$.

In the main text we are interested in performing T-duality transformations on the isometries of AdS$_3$. Previous results, \cite{Hassan:1999bv,Kelekci:2014ima} on the preservation of supersymmetry under U(1) and SU(2) T-dualities suggest that T-dualising on  AdS$_3$ will only lead to a solution that preserves supersymmety when its spinors are singlets with respect to the dualisation isometry. Thus any AdS$_3$ solution of Type II supergravity that preserves supersymmetry better be compatible with the spinors $\zeta_R$ for supersymmetry to be preserved after T-dualising on $\partial_r$. For SL(2) T-duality we have two options for dualisation isometry, we shall choose to dualise on SL(2)$_R$ as this contains the U(1) of the Hopf fiber. Thus far our discussion on spinors has been with respect to the frame \eqref{eq:AdS3frame}, note however that this is not the appropriate frame for SL(2) T-duality, for that the vielbein should also respect an SL(2) isometry. We shall take
\beq
e^{\underline{\mu}}_{3,\text{SL}(2)}= \frac{1}{2}(R^1,~R^2,~R^3)^{\underline{\mu}}.\label{eq:NATDframe}
\eeq
We could also use $L^i$ to span the frame, however we choose the right SL(2) to relate to the case of U(1) T-duality. The difference between the 2 frames is  simply a Lorentz transformation, specifically
\beq\label{eq:rotationtoSL2frame}
e^{\underline{\mu}}_{3,\text{SL}(2)}= {\cal R}^{\underline{\mu}}_{~\underline{\nu}}e^{\underline{\nu}}_{3},~~~~{\cal R} = \left(\begin{array}{ccc}1&0&0\\0&\cos t& s\sin t\\ 0&-s\sin t&\cos t\end{array}\right) \left(\begin{array}{ccc}\cosh x&0&s \sinh x\\0&1&0\\ s\sinh x&0&\cosh x\end{array}\right).
\eeq
Under such a transformation the spinors also transform as $\zeta \to  S \zeta$ where  $S$ solves the equation
\beq
S^{-1}\gamma^{\underline{\mu}}S= {\cal R}^{\underline{\mu}}_{~\underline{\nu}}  \gamma^{\underline{\nu}},
\eeq 
we find
\beq\label{eq:theSspinoraction}
S= e^{-\frac{s}{2}t i \sigma_2} e^{-\frac{s}{2}x\sigma_1}
\eeq
and so in the frame \eqref{eq:NATDframe} we have
\beq
\zeta_{R,\text{SL}(2)}= S \zeta_R=  \zeta_R^0,~~~~\zeta_{L,\text{SL}(2)}= S \zeta_L=e^{-\frac{s}{2}t i \sigma_2} e^{-\frac{s}{2}x\sigma_1}e^{-\frac{s}{2}r \sigma_3}\zeta_L^0=  G\zeta_L^0.
\eeq

\section{Deriving the SL(2) T-duality dual  fields}\label{sec:performingSL(2)T-duality}
In this appendix we shall derive the dual NS and RR fields that one gets after performing an SL(2) non-Abelian T-duality transformation on an AdS$_3$ solution of the form \eqref{eq:1.0}.  The procedure for doing this in the presence of non trivial RR fields was first worked out in  \cite{Sfetsos:2010uq} with a more detailed, and easier to follow, account given by \cite{Itsios:2013wd}. Both of these works focus on SU(2) T-duality, but the SL(2) case is rather analogous so we shall be brief here and refer the reader to \cite{Itsios:2013wd} for more details. The reader can also check  \cite{Ramirez:2021tkd} where SL(2) T-duality was applied to $\text{AdS}_3\times \text{S}^3\times \mathbb{T}^3$ and a briefer account on the content of this section, especially on the part regarding the transformation of the spinors, was also given.

The dual NS sector is computed via a non Abelian Buscher procedure: One first defines a sigma model with worldvolume coordinates $\sigma_{\pm}$ in terms of the NS fields with components along the SL(2) isometry directions. For a solution of the form \eqref{eq:1.0}, with trivial electric NS flux we have
\beq
ds^2=  g_{ij}R^i R^j+....,~~~~ B=....,~~~~ dB=H,~~~~ g_{ij}= \frac{e^{2A}}{4}\eta_{ij} 
\eeq
where ... have no components along the SL(2) directions and are thus spectators to the Buscher procedure. The Lagrangian of our sigma model is then simply
\beq
{\cal L}_0=  g_{ij} R^i_+ R^j_-,
\eeq
where $R^i_{\pm}=-2 s\text{Tr}\bigg[\tau^i \partial_{\pm} G G^{-1} \bigg]$ for $G\in$ SL(2) and $\tau_i$ the generators of $\mathfrak{sl}(2)$, consistent with the previous section.

The next step of the Buscher procedure is to rewrite  ${\cal L}_0$ in a locally equivalent form. For that one gauges the SL(2) isometry by introducing SL(2) gauge fields ${\cal A}_{\pm}= ({\cal A}_{\pm})^i \tau_i $ such that
\beq
\partial_{\pm}G\to D_{\pm}G = \partial_{\pm}G- {\cal A}_{\pm} G,
\eeq
but one additionally adds a Lagrange multiplier term to the action of the form $\text{Tr}(v {\cal F}_{\pm})$, where ${\cal F}_{\pm}$ is the non-Abelian field strength of ${\cal A}_{\pm}$ - this term setting ${\cal F}_{\pm}=0$ on shell. A dual sigma model is then generated if one integrates out the gauge fields. This theory generically depends on the 3 coordinates packaged within $G$ and the Lagrange multipliers $v_i$, and one needs to gauge fix such that only 3 of these remain. The simplest way to do this is to set
\beq
G= \mathbbm{1}.
\eeq  
The Lagrangian of the resulting dual sigma model is then given by 
\beq
\hat{{\cal L}}=  \partial_{-}v_i (M^{-1})^{ij}\partial_{-}v_j,~~~~ M_{ij}= g_{ij} + s f_{ij}^{~~k}v_k,
\eeq
where $f_{ij}^{~~k}$ are the structure constants of SL(2).  From this the dual metric and NS flux can be read off as
\beq
d\hat{s}^2=   dv_i(M^{-1})^{(ij)} dv_j+....,~~~~~ \hat B=\frac{1}{2}  (M^{-1})^{[ij]} dv_i\wedge dv_j+...
\eeq
where $...$ are the spectator parts of the original metric and NS 2-form. To compute the dual dilaton one needs to also consider quantum effects. These lead to 
\beq
e^{-\hat \Phi}=  c  e^{-\Phi}\sqrt{\det M},
\eeq
where $c$ is an arbitrary constant which can be set to any value by rescaling $g_s$.

The dual RR sector can be computed as in \cite{Sfetsos:2010uq}. One first observes that the dual sigma model defines a  pair of canonical frames. In the seed geometry, we had that the $d=10$ vielbein along the AdS$_3$ could be expressed in an SL(2) invariant way as
\beq
e^i =  \frac{e^{A}}{2} R^i,
\eeq
while the Buscher procedure defines 2 frames in the dual solutions
\beq
e^i_+= \frac{e^A}{2} (M^{-T})^{ij} dv_j,~~~~e^i_-=\frac{e^A}{2} (M^{-1})^{ij} dv_j.
\eeq
These are related by a Lorentz transformation
\beq
e_+ = \Lambda e_-,~~~~~ \Lambda =  M^{-T} M.
\eeq
As usual one can generate an action on spinors from the above $\epsilon\to \Omega\epsilon$ by solving the equation
\beq
\Omega^{-1}\Gamma^{\underline{M}} \Omega =  \Lambda^{\underline{M}}_{~\underline{N}} \Gamma^{\underline{N}}
\eeq
where $\Gamma^{\underline{M}} $ are a basis of $d=10$ flat index gamma matrices. The main contribution of 
 \cite{Sfetsos:2010uq} was to realise (through analogy with  \cite{Hassan:1999bv} ) that it is through the action of $\Omega$ that the dual RR sector is generated.
It is possible to show that $\Omega$ is given by
\beq
\Omega =  \frac{e^{3A}}{8\sqrt{\det M}} ( s \Gamma^{\underline{0}\underline{1}\underline{2}}+ \frac{4}{e^{2A}}v_i\Gamma^{\underline{i-1}}),\label{eq:Omegadef}
\eeq
where we have fixed an arbitrary sign. The dual RR fluxes can then be extracted from the following polyform identities
\beq
e^{\hat \Phi}\hat{F}=  t e^{\Phi} F \Omega^{-1},~~~~t^2=1,
\eeq
where $\Omega^{-1}$ acts on F through the Clifford map.

Let us now examine the symmetries of the dual solution. To this end we find it convenient to fix
\beq
v_3=-\frac{1}{2} s r \sinh x,~~~~ v_2=\frac{1}{2} r s \cosh x \sin t,~~~~v_3=\frac{1}{2}\cos t \cosh x,~~~~ c=4.
\eeq
The dual NS sector then becomes
\begin{align}
d\hat {s}^2&= \frac{e^{2A}}{4\Delta}ds^2(\text{AdS}_2)+ e^{-2A}dr^2+...,~~~~ \hat{B}= -\frac{s r}{2\Delta}\text{vol}(\text{AdS}_2)+....,~~~~e^{-\hat \Phi}= e^{-\Phi+A}r \sqrt{\Delta},\nn\\[2mm]
ds^2(\text{AdS}_2)&= -\cosh^2 xdt^2+dx^2,~~~~\text{vol}(\text{AdS}_2)= \cosh x dt \wedge dx.
\end{align}
So we see that the NS sector respects the SL(2) isometry of AdS$_2$. To compute the dual RR fluxes we use that
\beq
t e^{\Phi-\hat \Phi} \Omega^{-1}:~~ 1\to t( r -\frac{s e^{4A}}{8\Delta}\text{vol}(\text{AdS}_2))\wedge dr ,~~~~ e^{3A} \text{vol}(\text{AdS}_3)\to t\frac{e^{3A}}{2}(-s+ \frac{r}{2\Delta} \text{vol}(\text{AdS}_2)).
\eeq
We thus find the dual $d=10$ flux is
\beq
\hat F= t\bigg[f_{\pm}\wedge( r -\frac{s e^{4A}}{8\Delta}\text{vol}(\text{AdS}_2))\wedge dr\mp\frac{e^{3A}}{2}(-s+ \frac{r}{2\Delta} \text{vol}(\text{AdS}_2))\wedge \star_7 \lambda f_{\pm}\bigg],
\eeq
which also respects the isometries of AdS$_2$.

Finally before moving on to address the issue of supersymmetry let us give a frame that also respects the isometries of AdS$_2$, as $e^i_{\pm}$ do not.
Defining  $\hat e = \hat {\cal R} e_+$ for
\beq
\hat {\cal R}= \left(\begin{array}{ccc}\cosh Y&-\sinh Y&0\\-\sinh Y&\cosh Y&0\\0&0&1\end{array}\right) {\cal R}^{-1},~~~~ Y=\log\left(\frac{\Delta_-}{\Delta_+}\right),~~~~\Delta_{\pm}= 1\pm \frac{e^{2A}s}{2 r},
\eeq
where $\cal R$ is defined in \eqref{eq:rotationtoSL2frame},
we arrive at
\beq
\hat e^i=  (\frac{e^{A}}{2\sqrt{\Delta}} \cosh x dt,~ \frac{e^{A}}{2\sqrt{\Delta}} dx,~ e^{-A} dr)^i.
\eeq
The corresponding action on $d=10$ spinors (where $S$ is defined in \eqref{eq:theSspinoraction}) is 
\beq\label{eq:hatSaction}
\hat S=  e^{-\frac{1}{2}Y\Gamma_{\underline{0}\underline{1}}}S^{-1}\otimes \mathbbm{1}_{16},
\eeq
which will be useful in the next section. 

\section{Proving SL(2) T-duality preserves supersymmetry}\label{sec:appendixonSUSY}

In this appendix we shall prove that SL(2) T-duality preserves supersymmetry. We begin by deriving the reduced $d=7$ constraints for AdS$_3$ vacua, which will be necessary for proving that supersymmetry of the dual solutions is implied by that of the original solutions.

\subsection{Supersymmetry constraints for AdS$_3$ solutions}\label{eq:AdS3SUSYapppendix}
In this appendix we derive reduced $d=7$ spinorial conditions that  AdS$_3$ vacua of the form \eqref{eq:1.0} must obey to be supersymmetric. As in the main text we focus on those with purely magnetic NS 3-form, ie those for which $c_0=0$.\\ 

Given our choice of gamma matrices in \eqref{eq:10dgammas} the $d=10$ spinors should decompose as
\begin{align}
\epsilon_1 & = \zeta \otimes \theta_+ \otimes \chi_1, \quad \epsilon_2 = \zeta \otimes \theta_{\mp} \otimes \chi_2,\label{eq:AdS3spinors10d}
\end{align}
where $\zeta$ are real Killing spinors on AdS$_3$ obeying the relation
\begin{align}
\label{KSAdS3}
\nabla_{\mu}^{\text{AdS}_3} \zeta = \frac{s}{2} \gamma_{\mu} \zeta,~~~~ s^2=1,
\end{align}
and $\chi_{1,2}$ are independent Majorana spinors on $\text{M}_7$. The objects $\theta_{\pm}$ are auxiliary 2d vectors which are needed to make the dimensionality of the $d=10$ spinors correct when decomposed in terms of 3 and 7 dimensional ones. They are defined as
\begin{align} 
\theta_+ = \frac{1}{\sqrt{2}}
	\begin{pmatrix}
	1 \\
	- i
	\end{pmatrix}, \qquad
\theta_- & = \frac{1}{\sqrt{2}}
	\begin{pmatrix}
	1 \\
	 i
	\end{pmatrix},
\end{align}
such that $\epsilon_{1,2}$ are indeed Majorana-Weyl. The $\pm$ signs indicate $d=10$ chirality and so as before the upper/lower signs are taken in Type IIA/IIB. We also have the following identities which we will make much use of in the coming computations
\beq\label{eq:simplieifdentites}
\sigma_1 \theta_{\pm}  = \mp i \theta_{\mp},~~~~\sigma_3 \theta_{\pm}  = \theta_{\mp},~~~~\sigma_1 \sigma_3 \theta_{\pm} = \pm i \theta_{\pm}.
\eeq
Our task now is to insert our ansatz for the spinor into the necessary conditions for supersymmetry \eqref{eq:10dsusyeqs1}-\eqref{eq:10dsusyeqs4}. To achieve this we must first decompose the $d=10$ spin covariant derivative in terms of the factors that appear in the warped product. The curved space analogue of $\Gamma_{\underline{\mu}}$ in \eqref{eq:10dgammas} is
\beq
\Gamma_{\mu} =  e^{A} \gamma_{\mu}\otimes \sigma_3\otimes \mathbbm{1}_8,
\eeq 
as such we find that
\begin{align}
\label{AdS3covder}
\nabla_{\mu}^{(10)} &= \nabla_{\mu}^{\text{AdS}_3} +\frac{1}{2} \Gamma_{\mu}dA , \quad &\mu={0,1,2}, \nn \\
\nabla_{a}^{(10)} &= \nabla_{a}, \quad & a={3,..,9},
\end{align}
where $\nabla_{a}$ is the spin covariant derivative on M$_7$. 
Making use of \eqref{KSAdS3} and \eqref{eq:simplieifdentites} and the identities
\beq
F= (\mathbb{I}_{32}+\hat\Gamma)f_{\pm},~~~~\lambda(F)= (\mathbb{I}_{32}\mp \hat\Gamma)\lambda(f_{\pm}),
\eeq
which follow from \eqref{eq:usefulid}, we find that \eqref{eq:10dsusyeqs1}-\eqref{eq:10dsusyeqs4} reduce to the following $d=7$ conditions
\begin{subequations}
\begin{align}
& (s e^{-A}- i  dA ) \chi_1 + \frac{1}{4} e^{\Phi } \beta_{\pm} f_{\pm} \chi_2 =0, \label{seedsolutionsusyeqs1} \\[2mm]
& (s e^{-A}  \pm idA ) \chi_2 +  \frac{1}{4} e^{\Phi } \beta_{\pm}^* \lambda(f_{\pm}) \chi_1 =0, \label{seedsolutionsusyeqs2}\\
& (\nabla_a - \frac{1}{4} H_{a}) \chi_1 + \frac{1}{8} e^{\Phi} i \beta_{\pm}^* f_{\pm} \gamma_a \chi_2 =0,\label{seedsolutionsusyeqs3}  \\[2mm]
&  (\nabla_a + \frac{1}{4} H_{a}) \chi_2 - \frac{1}{8} e^{\Phi} i \beta_{\pm}^* \lambda(f_{\pm}) \gamma_a \chi_1 =0,\label{seedsolutionsusyeqs4}  \\[2mm]
& \left( \frac{3}{2} s e^{-A} -i (\frac{3}{2}  dA +\nabla + \frac{i}{4} H - d \Phi )\right)\chi_1=0, \label{seedsolutionsusyeqs5} \\[2mm]
& \left( \frac{3}{2} s e^{-A} \pm i(\frac{3}{2}  dA + \nabla + \frac{1}{4} H -  d \Phi )\right)\chi_2=0\label{seedsolutionsusyeqs6}. 
\end{align}
\end{subequations}
where $\beta_{\pm}$ is a constant, specifically
\begin{align}
\beta_{+} = & 1 , \quad \beta_- = i.
\end{align}

\subsection{U(1) T-duality on the Hopf fiber of AdS$_3$}
In this appendix we show that U(1) T-duality on the Hopf fiber of AdS$_3$ preserves supersymmetry.

As first established in \cite{Hassan:1999bv}, U(1) T-duality preserves the portion of the supercharges in the orginal solution that are not charged under the U(1). When one performs a Hopf fiber T-duality on AdS$_3$ then, we should identify $\zeta$ in \eqref{eq:AdS3spinors10d} with the SL(2)$_R$ invariant Killing spinor $\zeta_R$ in \eqref {eq:AdS3spinors}. Given an AdS$_3$ solution of the form \eqref{eq:1.0}, in the conventions of \eqref{sec:detailsonAdS3} the result of T-dualising on the Hopf fiber isometry $\partial_r$ is the following 
\begin{align}
d\hat{s}^2&= \frac{e^{2A}}{4} ds^2(\text{AdS}_2)+ e^{-2A} dr^2+ ds^2(\text{M}_7),~~~~e^{-\hat \Phi}= e^{-\Phi+A},\nn\\[2mm]
\hat H&=  -\frac{s}{2}r\text{vol}(\text{AdS}_2)+H,~~~~\hat F=t (f_{\pm}\wedge dr\mp \frac{1}{4}e^{3A}\text{vol}(\text{AdS}_2)\wedge \star_7 \lambda f_{\pm}), 
\end{align}
where again $s^2=t^2=1$.
Here the dual RR flux is generated via  $e^{\hat \Phi}\hat F=  e^{\Phi} F \Gamma_{\underline{r}}$ where we have split the $\mu$ index as $\mu=(\mu_2,r)$ where $\mu_2=0,1$ and runs over the AdS$_2$ directions.

A main result of \cite{Hassan:1999bv} is the derivation of the precise form of the dual Killing spinors in terms of the original one, when U(1) T-duality is performed. The map is
\beq\label{eq:dualu1spinors}
\epsilon_1\to \hat \epsilon_1= \epsilon_1,~~~~\epsilon_2\to \hat \epsilon_2= \tilde{t}\Gamma_{\underline{r}}\epsilon_2,~~~\tilde{t}^2=1,
\eeq
where we shall fix $\tilde{t}$ momentarily. For this to hold we must be in a canonical frame in which the vielbein decomposes as a U(1) fibration over a 9-d base, ie we take the $d=10$ vielbein in the original solution to be
\beq
e^{\underline{\mu}}_{10}= e^{A} e^{\underline{\mu}}_3, ~~~~ e^{\underline{a}}_{10}= e^{\underline{a}}
\eeq
where $e^{\underline{\mu}}_3$ is defined in \eqref{eq:AdS3frame} and $e^a$ is a vielbein on $\text{M}_7$. The dual frame on the other hand is
\beq
\hat e^{\underline{\mu}_2}_{10}= \frac{e^A}{2} e^{\underline{\mu}_2}_2,~~~~\hat e^{\underline{r}}_{10}= e^{-A} dr,~~~~ \hat e^{\underline{a}}_{10}=e^{\underline{a}}
\eeq
where $e^{\underline{\mu}}_2$ is a vielbein on unit radius AdS$_2$ (ie one can take $e^{\underline{\mu}_2}_2= (\cosh x dt,dx)^{\underline{\mu}_2}$).

Our task is once more to derive the reduced $d=7$ conditions that follow from considering \eqref{eq:10dsusyeqs1}-\eqref{eq:10dsusyeqs4} on the dual solution - note that T-duality maps a solution in IIA to one in IIB so one needs to be careful with the signs.  We find that the $d=10$ covariant derivative now decomposes as
\begin{align}
\label{AdS2covderATD}
\nabla_{\mu_2}^{(10)} &= \nabla_{\mu_2}^{\text{AdS}_2} + \frac{1}{2}  \Gamma_{\mu_2}dA,~~~~
\nabla_{r}^{(10)} = \partial_{r} - \frac{1}{2} \Gamma_{r}dA ,~~~~
\nabla_{a}^{(10)} = \nabla_{a}.
\end{align}
The dual $d=10$ spinors decompose as
\beq
\hat \epsilon_1=\zeta\otimes \theta_+\otimes \chi_1,~~~~~\hat \epsilon_2=  \Gamma_{\underline{r}}\epsilon_2= \gamma_{\underline{r}}\zeta\otimes \theta_{\pm}\otimes \chi_2,
\eeq
if we then consider \eqref{eq:10dsusyeqs3} we find it decomposes as
\beq
2 e^{-A}\gamma^{\mu_2}\nabla^{\text{AdS}_2}_{\mu_2}\zeta\otimes \theta_-\otimes \chi_1-\zeta\otimes \theta_-\otimes \bigg[\frac{s}{2}e^{-A}+i(\nabla+\frac{3}{2}dA-\frac{1}{4}H-d\Phi)\bigg]\chi_1=0,
\eeq
which reproduces \eqref{seedsolutionsusyeqs5} if 
\beq
\nabla_{\mu_2}^{\text{AdS}_2}\zeta= \frac{s}{2} \gamma_{\mu_2}\zeta,\label{eq:AdS2cond}
\eeq
but this is precisely what the SL(2)$_R$ singlet spinor on AdS$_3$ does obey. If we now consider  \eqref{eq:10dsusyeqs1} along the internal directions we find it reduces to
\beq
(\nabla_a-\frac{1}{4}H_a)\epsilon_1- t \tilde t\frac{e^{\Phi}}{16} F \Gamma_a \epsilon_2=0
\eeq 
which reproduces  \eqref{seedsolutionsusyeqs3} if we fix
\beq
t\tilde{t}=-1.
\eeq 
Given this one finds that inserting the dual solution into the rest of  \eqref{eq:10dsusyeqs1}-\eqref{eq:10dsusyeqs4} reproduces all of \eqref{seedsolutionsusyeqs1}-\eqref{seedsolutionsusyeqs6}. The key point however is that only one of the 2 types of Killing spinors that AdS$_3$ supports is preserved by the duality, what we referred to as $\zeta_R$ in appendix \ref{sec:detailsonAdS3} - the other is always projected out, ie what we earlier called $\zeta_L$ does not obey \eqref{eq:AdS2cond}.

\subsection{SL(2) T-duality on  AdS$_3$}

In this appendix we show that SL(2) T-duality performed on the SL(2)$_R$ isometry of  AdS$_3$ preserves supersymmetry.

By analogy with the previous appendix and the results of \cite{Kelekci:2014ima} for SU(2) T-duality, we make the following ansatz for how SL(2) T-duality acts on the spinors of an AdS$_3$ solution
\beq
\hat \epsilon_1=\epsilon_1,~~~~\hat \epsilon_2=\tilde{t}\Omega\epsilon_2,\label{eq:canonicalNATDspinoransatz}
\eeq
where $\Omega$ is defined in \eqref{eq:Omegadef}. Like with U(1) T-duality, SL(2)$_R$ T-duality has a canonical frame, namely the vielbein of the original solution 
should take the form
\beq
e^{\underline{\mu}}_{10}= e^{A} e^{\underline{\mu}}_{3,SL(2)}, ~~~~ e^{\underline{a}}_{10}= e^{\underline{a}},\label{eq:seedNATD}
\eeq
where $e^{\underline{\mu}}_{3,SL(2)}$ is defined in \eqref{eq:NATDframe}. Importantly this differs from   the frame of U(1) T-duality in the previous appendix by a Lorentz transformation as described in appendix \ref{sec:detailsonAdS3}. This means that if we take our original Killing spinors to be \eqref{eq:AdS3spinors10d} and identify $\zeta=\zeta_R$ in the U(1) T-duality frame, they are now given by
\beq
\epsilon_1= S\zeta\otimes \theta_{+}\otimes \chi_1,~~~~\epsilon_2=S\zeta\otimes \theta_{\mp}\otimes \chi_2
\eeq
in the frame \eqref{eq:seedNATD}. Likewise, as explained towards the end of appendix \ref{sec:performingSL(2)T-duality}, in the canonical frame of SL(2) T-duality the isometries of AdS$_2$ are not manifest, to reach such a frame one needs to do another  Lorentz tranformation that will also act on \eqref{eq:canonicalNATDspinoransatz}. A dual frame that respects the isometries of AdS$_2$ has the $d=10$ vielbein
\beq
\hat e^{\underline{\mu}_2}_{10}= \frac{e^A}{2 \sqrt{\Delta}} e^{\underline{\mu}_2}_2,~~~~\hat e^{\underline{r}}_{10}= e^{-A} dr,~~~~ \hat e^{\underline{a}}_{10}=e^{\underline{a}},\label{eq:NATDads2frame}
\eeq
where we remind the reader of the following functions
\beq
\Delta=  1-\frac{e^{4A}}{4 r^2},~~~~~\Delta_{\pm}= 1\pm \frac{e^{2A} s}{2 r}
\eeq
The combined actions of the Lorentz transformations required to reach this frame mean that our dual spinors now take the form
\begin{align}
\hat{\epsilon}_1&= (N_++N_-\gamma_{\underline{r}})\zeta\otimes \theta_{+}\otimes \chi_1,~~~~\hat{\epsilon}_2=\tilde{t}\Gamma_{\underline{r}}(N_+-N_-\gamma_{\underline{r}})\zeta\otimes \theta_{\mp}\otimes \chi_2,\nn\\[2mm]
N_{\pm}&=\frac{1}{2\Delta^{\frac{1}{4}}}(\sqrt{\Delta_+}\pm \sqrt{\Delta_-}),
\end{align}
where the action of \eqref{eq:hatSaction} is what maps the spinor between the frame of \eqref{eq:seedNATD} and \eqref{eq:NATDads2frame}, the respective $S$ and $S^{-1}$ terms canceling in the process. The terms appearing in the $d=10$ supersymmetry conditions can also be written in terms of $\Delta,\Delta_{\pm}$, after applying the Clifford map these become
\begin{align}
\hat{H}&= H - 2 s e^{-A} \Gamma^{\underline{0}\underline{1}\underline{r}} + \frac{2 }{ \Delta_+ - \Delta_-} d\log(\Delta) \Gamma^{\underline{0}\underline{1}}, \nn \\[2mm]
d\hat{\Phi} &= d\Phi - dA - \frac{1}{2} d\log(\Delta) - s (\Delta_+ - \Delta_-) e^{-A} \Gamma^{\underline{r}}, \nn \\[2mm]
 e^{\hat{\Phi}} \hat{F} &= t \frac{e^{\Phi} }{\sqrt{\Delta}} (1 + \Gamma) \left( f_{\pm} \Gamma^{\underline{r}} -   \frac{1}{2} (\Delta_+ - \Delta_-)f_{\pm} \Gamma^{\underline{0}\underline{1}\underline{r}}\right). \nn \\[2mm]
 e^{\hat{\Phi}} \lambda(\hat{F}) &= t \frac{e^{\Phi} }{\sqrt{\Delta}} (1 \pm \Gamma) \left(  \Gamma^{\underline{r}}  + \frac{1}{2} (\Delta_+ - \Delta_-)  \Gamma^{\underline{0}\underline{1}\underline{r}} \right) \lambda(f_{\pm}) .
\end{align}
In this case the the spin covariant derivative decomposes as
\beq
\nabla_{\mu_2}^{(10)} = \nabla_{\mu_2}^{\text{AdS}_2} + \frac{1}{2}\Gamma_{\mu_2} dA  - \frac{1}{4} \Gamma_{\mu_2} d(\log(\Delta)),~~~~
\nabla_{r}^{(10)} = \partial_r - \frac{1}{2} \Gamma_{r}dA,~~~~ \nabla_{a}^{(10)} = \nabla_{a} \label{AdS2covderNATD}.
\eeq
We also have the following useful identities
\begin{align}
\label{propertiesNpm}
& d(N_+ \pm N_- \gamma_{\underline{r}}) = \mp \frac{1}{ 2 (\Delta_+ - \Delta_-)} d\log(\Delta) \gamma_{\underline{r}} (N_+ \pm N_- \gamma_{\underline{r}}), \nn \\[2mm]
& 2 N_{\pm}^2 = \frac{1}{\sqrt{\Delta}} \pm 1    , \qquad  N_+ N_- = \frac{\Delta_+ - \Delta_-}{4 \sqrt{\Delta}}.
\end{align}

We can now proceed with reducing the $d=10$ Killing spinor equations as before. We first consider \eqref{eq:10dsusyeqs3} and find that the terms involving $\log(\Delta)$ and $\partial_r$ mutually cancel leaving
\begin{align}
& \bigg( \Gamma^{\mu_2} \nabla_{\mu_2}^{\text{AdS}_2}  + \frac{3}{2} dA  + \nabla^{\chi_1} - \frac{1}{4} H + \frac{s}{2} e^{-A} \Gamma^{\underline{0}\underline{1}\underline{r}} - d\Phi  +  s (\Delta_+ - \Delta_-) e^{-A} \Gamma^{\underline{r}} \bigg) \hat{\epsilon}^1=0,\label{eq:natdsusy1}
\end{align}
where $\nabla^{\chi_1}$ indicates that $\nabla$ only acts on $\chi_1$ in this expression, not $N_{\pm}$. Through \eqref{propertiesNpm} it is possible to establish that
\begin{align}
\left( (N_+ - N_- \gamma_{\underline{r}} ) 2 \sqrt{\Delta} s e^{-A} + s (\Delta_+ - \Delta_-) \gamma_{\underline{r}} (N_+ + N_- \gamma_{\underline{r}} ) \right) = 2 s e^{-A} (N_+ + N_- \gamma_{\underline{r}} ),
\end{align}
so that \eqref{eq:natdsusy1} decomposes as
\begin{align}
&2\sqrt{\Delta}e^{-A}(N_+ - N_- \gamma_{\underline{r}} ) (\gamma^{\mu_2}\nabla^{\text{AdS}_2}_{\mu_2}- s)\zeta\otimes\theta_-\otimes \chi_1\nn\\[2mm] 
&+(N_+ + N_- \gamma_{\underline{r}} ) \zeta \otimes \theta_- \otimes\left( \frac{3}{2} s e^{-A} - i (\frac{3}{2} dA + \nabla - \frac{1}{4} H - d \Phi) \right)\chi_1=0,
\end{align}
where  the first line vanishes if $\zeta=\zeta_R$ and the second reproduces \eqref{seedsolutionsusyeqs5}. Assuming from now that indeed $\zeta=\zeta_R$, so that \eqref{eq:AdS2cond} holds, with an analogous computation we find that \eqref{eq:10dsusyeqs4} reduces to 
\begin{align}
\gamma_{\underline{r}} (N_+ - \gamma_{\underline{r}} N_-) \zeta \otimes \theta_{\mp} \otimes  \left( \frac{3}{2} s e^{-A} \pm i ( \frac{3}{2} dA + \nabla + \frac{1}{4} H - d \Phi ) \right)\chi_2 =0, 
\end{align}
which likewise reproduces \eqref{seedsolutionsusyeqs6}.

Turning our attention now to the part of \eqref{eq:10dsusyeqs1} along the internal directions we find, after extensive use of the identities in \eqref{propertiesNpm}, that  it reduces to
\begin{align}
\left( N_+ + N_- \gamma_{\underline{r}}\right) \zeta \otimes \theta_+  \left( \left( \nabla_a -\frac{1}{4} H_a \right)\chi_1 - t \tilde{t} \frac{e^{\Phi} }{8} i \beta_{\pm}^*  f_{\pm} \gamma_a \chi_2 \right) =0,
\end{align}
which like the U(1) case reproduces  \eqref{seedsolutionsusyeqs3} if we fix
\beq
t\tilde{t}=-1.
\eeq 
Given this the remaining components of \eqref{eq:10dsusyeqs1}, along respectively $r$ and $\mu_2$ can be massaged to
\begin{align}
&\Gamma_{\underline{r}} \left( N_+ + N_- \gamma_{\underline{r}}\right) \zeta \otimes \theta _+ \otimes \left( \left( e^{-A}s - i  dA \right) \chi_1+\frac{e^{\Phi }}{4 } \beta_{\pm} f_{\pm} \chi_2 \right) =0,\\[2mm]
&\Gamma_{\mu_2}\left( N_+ (2- \sqrt{\Delta}) - \gamma_{\underline{r}}N_- (2 + \sqrt{\Delta})\right) \zeta \otimes \theta_- \otimes  \left( \left( e^{-A}s  - i  dA \right) \chi_1 + \frac{1 }{4} e^{\Phi } \beta_{\pm} f_{\pm} \chi_2 \right)=0,\nn
\end{align}
which both reproduce \eqref{seedsolutionsusyeqs1}. Note that simplifying the second of these is more challenging and the following identity is useful
\begin{align}
d\log(\Delta) = \frac{\Delta-1}{\Delta} \left(4 dA - 2s e^{-A} (\Delta_+ - \Delta_-)\Gamma^{\underline{r}} \right).
\end{align}
Finally we consider \eqref{eq:10dsusyeqs2}. Through similar means to before it is possible to reduce this along each of $a,r,\mu_2$ to the form
\begin{align}
&  (N_+ - \gamma_{\underline{r}} N_-) \zeta \otimes \theta_{\pm} \otimes \left( \left( \nabla_a + \frac{1}{4} H_a \right) \chi_2 - \frac{e^{\Phi}}{8} i \beta_{\pm}^* \lambda(f_{\pm}) \gamma_a \chi_1 \right)=0,\nn\\[2mm]
& \Gamma_{r}(N_+ - \gamma_{\underline{r}} N_-) \zeta \otimes \theta_{\pm} \otimes \left( (s e^{-A} \pm i dA ) \chi^2 +  \frac{1}{4} e^{\Phi} \beta_{\pm}^* \lambda(f_{\pm}) \chi_1 \right) = 0, \nn\\[2mm] 
& \Gamma_{\mu_2}(N_+ - \gamma_{\underline{r}} N_-) \zeta \otimes \theta_{\pm} \otimes \left( (s e^{-A} \pm i dA ) \chi^2 +  \frac{1}{4} e^{\Phi} \beta_{\pm}^* \lambda(f_{\pm}) \chi_1 \right) = 0, 
\end{align}
which reproduce \eqref{seedsolutionsusyeqs2} and \eqref{seedsolutionsusyeqs4}.

We have thus established that performing an SL(2)$_R$ T-duality on a supersymmetric AdS$_3$ solution that supports spinors invariant under SL(2)$_R$ preserves supersymmetry. Just as with the U(1) case any spinors charged under SL(2)$_R$ will be projected out.

\section{Details of the quiver construction}\label{fieldtheory}

In this appendix we present some details of the construction of the 1d quivers discussed in section \ref{field-theory}.

As discussed in the main text, our Hanany-Witten brane set-up consists on D-branes stretched between D(p+2)-branes, with orthogonal NS5-branes lying between them, instead of the more conventional scenario in which the Dp-branes are stretched between NS5-branes, with orthogonal D(p+2)-branes between them, which allows to directly read the quiver field theory. In this situation in order to obtain the quiver field theory we need to perform Hanany-Witten moves that produce D-branes stretched between NS5-branes with D(p+2)-branes orthogonal to them. We do this using  that the D5-D7-NS5 and D1-D3-NS5 brane subsystems in our brane set-up are related by T-duality to the D3-D5-NS5 brane intersection studied in  \cite{Hanany:1996ie}, and also to the D2-D4-NS5 two dimensional brane intersection underlying the $\text{AdS}_3\times \text{S}^3\times \mathbb{T}^3\times {\cal I}$ solutions to Type IIA constructed in \cite{Lozano:2022ouq}, from  which our solutions are derived through SL(2) T-duality. We will then follow closely \cite{Lozano:2022ouq} where the results in \cite{Hanany:1996ie} were used to describe the 2d quiver field theory living in this brane intersection. In this appendix we basically  rewrite the analysis therein particularised to the D5-D7-NS5 and D1-D3-NS5 subsystems of the brane set-up depicted in Table \ref{D1D7NS5D5D3F1}.

We start discussing the D5-D7-NS5 brane subsystem. Following   \cite{Lozano:2022ouq} we use the following definitions for the linking numbers
\begin{eqnarray}
&&l_i=n_i+L_i^{NS5}, \qquad \text{for the D7-branes} \nonumber\\
&&\hat{l}_j=-\hat{n}_j+R_j^{D7}, \qquad \text{for the NS5-branes}, \nonumber
\end{eqnarray}
where $n_i$ is the number of D5-branes ending on the $i$th D7-brane from the right minus the number of D5-branes ending on it from the left, $\hat{n}_j$ is the same quantity for the $j$th NS5-brane, $L_i^{NS5}$ is the number of NS5-branes lying on the left of the $i$th D7-brane, and $R_j^{D7}$ is the number of D7-branes lying on the right of the $j$th NS5-brane. 

We proceed by computing the number of D5-branes that end on the left of a collection of D7-branes and on the right of a collection of NS5-branes, given by $N=\sum_{i=1}^p l_i=\sum_{j=1}^{\hat{p}}\hat{l}_j$, with $p$ and $\hat{p}$ the numbers of D7 and NS5 branes. Here the partition $N=\sum_{j=1}^{\hat{p}}\hat{l}_j$ has to be such that $\hat{l}_1\ge \hat{l}_2\ge \dots \hat{l}_{\hat{p}}$. To read the quiver we consider this partition plus the partition $N=\sum_{s=1}^r M_s q_s$, constructed from a list of positive integers satisfying $q_1\ge q_2\ge \dots q_r$, chosen such that the number of terms in the decomposition that are equal or bigger than a given integer $j$, that we denote as $m_j$, satisfy that
\begin{equation}\label{mj}
\sum_{j=1}^{i}m_j\ge \sum_{j=1}^{i}\hat{l}_j, \quad \text{for}\, \,\,  i=1,\dots, \hat{p}.
\end{equation}
$M_s$ is then the number of times each integer $q_s$ appears in the partition. 
From these two partitions the ranks of the gauge groups of the quiver are computed as
\begin{equation}
N_i=\sum_{j=1}^i (m_j-\hat{l}_j).
\end{equation}
In turn, $M_s$ give the ranks of the fundamental matter groups that couple to each of the gauge groups. A very detailed account of this construction can be found in \cite{Assel:2011xz}. 

\begin{figure}
\centering
\includegraphics[scale=0.55]{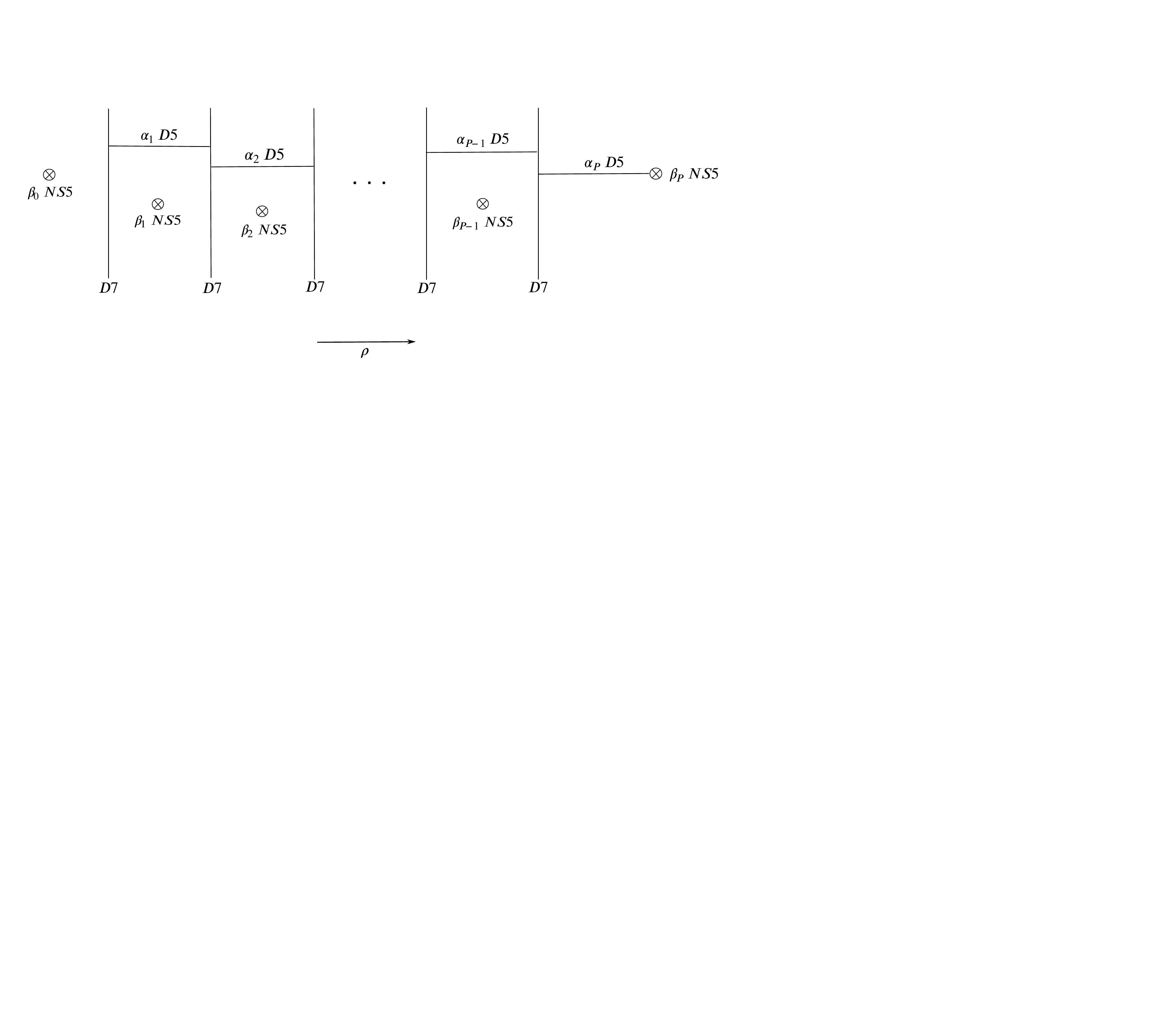}
\vspace{-15cm}
\caption{Brane set-up for the D5-D7-NS5 branes along the $\rho$ direction, for $r$ constant, in units of $q$.}
\label{brane-set-up-D5D7}
\end{figure}

We now apply this procedure to the brane set-up associated to our solutions, that we terminate with $\beta_P$ anti NS5-branes at the end of the space, at $\rho=P+1$. For the D5-D7-NS5 subsystem the resulting brane set-up is the one depicted in Figure \ref{brane-set-up-D5D7}. 
From this configuration the linking numbers are read exactly as in \cite{Lozano:2022ouq}, substituting the D2 for D5-branes and the D4 for D7-branes. Doing this, we find for the D7-branes
\begin{equation}
l_i=\sum_{r=0}^{i-2}\beta_r+2\beta_{i-1}, \quad i=1,\dots, P
\end{equation}
and for the NS5-branes
\begin{eqnarray}
&&\hat{l}_1=\hat{l}_2=\dots =\hat{l}_{\beta_0}=P,\nonumber\\
&&\hat{l}_{\beta_0+1}=\hat{l}_{\beta_0+2}=\dots =\hat{l}_{\beta_0+\beta_1}=P-1,\nonumber\\
&& \hspace{4cm}\vdots\nonumber\\
&&\hat{l}_{\beta_0+\beta_1+\dots +\beta_{P-3}+1}=\hat{l}_{\beta_0+\beta_1+\dots +\beta_{P-3}+2}=\dots =
\hat{l}_{\beta_0+\beta_1+\dots +\beta_{P-2}}=2,\nonumber\\
&&\hat{l}_{\beta_0+\beta_1+\dots +\beta_{P-2}+1}=\dots =\hat{l}_{\beta_0+\beta_1+\dots +\beta_{P-1}}=1,\nonumber\\
&&\hat{l}_{\beta_0+\beta_1+\dots +\beta_{P-1}+1}=\dots =\hat{l}_{\beta_0+\beta_1+\dots +\beta_{P-1}+\beta_P}=1.
\end{eqnarray}
From the linking numbers we construct the total number of D5-branes ending on D7-branes on the left and NS5-branes on the right. This is given by
\begin{equation}
N=\sum_{i=1}^P l_i=\sum_{j=1}^{\beta_0+\dots +\beta_P}\hat{l}_j=\sum_{k=0}^{P-1} (P-k+1)\beta_{k}.
\end{equation}
Now, from $N$ we define the two partitions that will allow us to read the quiver CFT. The NS5-branes in our brane set-up are ordered such that $\hat{l}_1\ge \hat{l}_2\ge \dots \ge \hat{l}_{\hat{\beta_0+\dots +\beta_P}}$. These linking numbers define then one of the two partitions, $N=\sum_{j=1}^{\beta_0+\dots +\beta_P}\hat{l}_j$.
In turn, for the D7-branes we take 
\begin{equation} \label{partition}
N=\underbrace{\beta_0}+\underbrace{\beta_0+\beta_1}+\underbrace{\beta_0+\beta_1+\beta_2}+\dots +\underbrace{\beta_0+\beta_1+\dots +\beta_{P-2}}+2\underbrace{(\beta_0+\beta_1+\dots +\beta_{P-1})}
\end{equation}
from which we find
\begin{eqnarray}
&&m_1=m_2=\dots =m_{\beta_0}=P+1, \nonumber\\
&&m_{\beta_0+1}=\dots=m_{\beta_0+\beta_1}=P, \nonumber\\
&& \hspace{4cm}\vdots\nonumber\\
&&m_{\beta_0+\beta_1+\dots +\beta_{P-3}+1}=\dots =m_{\beta_0+\beta_1+\dots +\beta_{P-2}}=3, \nonumber\\
&&m_{\beta_0+\beta_1+\dots +\beta_{P-2}+1}=\dots =m_{\beta_0+\beta_1+\dots +\beta_{P-1}}=2.
\end{eqnarray}
These numbers satisfy the condition \eqref{mj}  $\forall i=1,\dots, (\beta_0+\dots +\beta_P)$.
We then find for the ranks of the gauge groups
\begin{eqnarray}
&&N_1=m_1-\hat{l}_1=P+1-P=1, \quad N_2=N_1+m_2-\hat{l}_2=2, \quad \dots \quad N_{\beta_0}=\beta_0, \nonumber\\
&&N_{\beta_0+1}=\beta_0+1, \quad \dots \quad 
N_{\beta_0+\beta_1+\dots + \beta_{P-1}}= \beta_0+\beta_1+\dots + \beta_{P-1}, 
\end{eqnarray}
to then start decreasing
\begin{equation}
N_{\beta_0+\beta_1+\dots \beta_{P-1}+1}= \beta_0+\beta_1+\dots + \beta_{P-1}-1, \quad \dots \quad ,
N_{\beta_0+\beta_1+\dots \beta_{P-1}+\beta_P-1}=1.
\end{equation}
That is, the ranks of the gauge groups increase in units of 1 till the value $\beta_0+\beta_1+\dots +\beta_{P-1}$ is reached, to then start decreasing, again in units of one, till the gauge group of rank 1 is reached, corresponding to the D5-branes stretched between the last pair of NS5-branes. 

\begin{figure}
\centering
\includegraphics[scale=0.65]{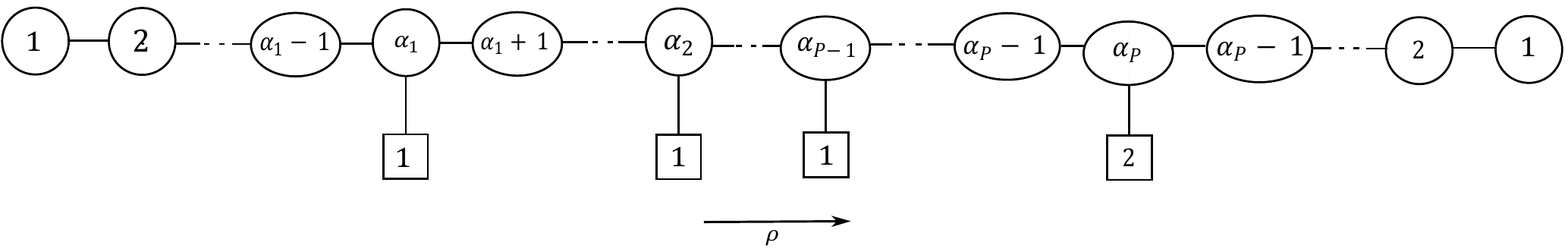}
\vspace{-17cm}
\caption{Quiver associated to the D5-D7-NS5 brane subsystem for $r$ constant, written in terms of 2d (4,4) multiplets. Circles denote (4,4) vector multiplets and black lines (4,4) bifundamental hypermultiplets. The gauge groups with ranks $\alpha_k$, with $k=1,\dots, P-1$ to the left of the gauge group with rank $\alpha_P$ have U(1) flavour symmetries. The gauge group with rank $\alpha_P$ has U(2) flavour symmetry. The rest of gauge groups do not have attached any flavours. We have taken units of $q$.}
\label{quiverT3_1}
\end{figure}

Finally, from the partition \eqref{partition} we have that
\begin{equation}\label{flavourgroups}
M_{\beta_0}=M_{\beta_0+\beta_1}=\dots = M_{\beta_0+\beta_1+\dots +\beta_{P-2}}=1,\quad M_{\beta_0+\beta_1+\dots +\beta_{P-1}}=2.
\end{equation}
This implies that the gauge groups with ranks $\beta_0=\alpha_1$, $\beta_0+\beta_1=\alpha_2$, till $\beta_0+\dots + \beta_{P-2}=\alpha_{P-1}$ have U(1) flavour groups, while the gauge group with rank $\beta_0+\beta_1\dots +\beta_{P-1}=\alpha_P$ has flavour group U(2). The rest of gauge groups have no flavour groups attached. The resulting quiver is depicted in Figure \ref{quiverT3_1}.  
In this quiver we have decomposed the 5d $\mathcal{N}=1$ vector multiplets and hypermultiplets in terms of 1d (4,4) multiplets, which are the appropriate ones to describe the D1-D3-NS5 subsystem and therefore the final quantum mechanics that arises in our brane intersection. Indeed, it is clear that the quiver associated to the D1-D3-NS5 brane subsystem of the brane set-up is obtained from the one describing the D5-D7-NS5 subsystem scaling it by $p=n+\frac12$, given the D1 and D3 brane quantised charges found in \eqref{QD1intervals} and \eqref{QD3intervals}. This quiver is depicted in Figure \ref{quiverD1D3} in subsection \ref{quiver}.  Therefore, our notation for both quivers is that circles denote (4,4) vector multiplets (coming from the reduction of 5d $\mathcal{N}=1$ vector multiplets for the D5-D7-NS5 subsystem) and black lines (4,4) bifundamental hypermultiplets (coming from the reduction of 5d $\mathcal{N}=1$ bifundamentals for the D5-D7-NS5 subsystem). These massless modes arise from the quantisation of the open strings stretched between the D1-branes (or wrapped D5-branes) and between the D1 and the D3 branes (or wrapped D5 and D7 branes), as follows:
\begin{itemize}
\item D1-D1 strings: There are two cases to consider. Open strings with both ends on the same stack of D1-branes between NS5-branes give rise to a (4,4) vector multiplet, composed of a (0,4) vector multiplet plus a (0,4) adjoint twisted hypermultiplet coming from the motion of the D1-branes along the $(x^6,x^7,x^8,x^9)$ directions. Since these scalars are charged under the R-symmetry they combine into a twisted hypermultiplet. In turn, the strings with end points on adjacent stacks of D1-branes separated by an NS5-brane contribute with a (4,4) bifundamental hypermultiplet, since the intersection with the NS5-branes fixes the degrees of freedom along the $(x^6,x^7,x^8,x^9)$ directions, leaving behind the scalars associated to the directions along the torus plus the scalar arising from the KK reduction of the 2d gauge field along the field theory direction, all of them uncharged under the R-symmetry. 

\item D1-D3 strings: Strings with one end on D1-branes and the other on D3-branes in the same interval between NS5-branes contribute with fundamental (4,4) hypermultiplets, associated to the motion of the string along the torus plus the component of the 2d gauge field along the field theory direction. This is related to the set-up in 
\cite{Hanany:1996ie} by T-duality, with the (4,4) hypermultiplet arising as the reduction to 1d of the 3d $\mathcal{N}=4$ hypermultiplet.

\end{itemize}

\end{document}